\documentclass[preprint,12pt]{elsarticle}

\usepackage{epsfig,graphics,amssymb,amsmath,subeqnarray,wasysym,mathtools}
\usepackage{subfigure}
\usepackage{xcolor}
\usepackage{pstricks}

\journal{Journal of Computational Physics}

\begin{document}

\begin{frontmatter}

\title{Computation of eigenvalue sensitivity to base flow modifications in a discrete framework: Application to open-loop control}

\author[rvt]{Cl\'ement Mettot}
\ead{clement.mettot@onera.fr}

\author[rvt]{Florent Renac}
\ead{florent.renac@onera.fr}

\author[rvt]{Denis Sipp} 
\ead{denis.sipp@onera.fr}

\address[rvt]{ONERA-The French Aerospace Lab, 29 avenue de la Division Leclerc, 92320 Ch\^atillon, France}

\begin{abstract}
A fully discrete formalism is introduced to perform stability analysis of a turbulent compressible flow whom dynamics is modeled with the Reynolds-Averaged Navier-Stokes (RANS) equations. The discrete equations are linearized using finite differences and the Jacobian is computed using repeated evaluation of the residuals. Stability of the flow is assessed solving an eigenvalue problem. The sensitivity gradients which indicate regions of the flow where a passive control device could stabilize the unstable eigenvalues are defined within this fully discrete framework. Second order finite differences are applied to the discrete residual to compute the gradients. In particular, the sensitivity gradients are shown to be linked to the Hessian of the RANS equations. The introduced formalism and linearization method are generic:  the code used to evaluate the residual of the RANS equations can be used in a black box manner, and the complex linearization of the Hessian is avoided. The method is tested on a two dimensional deep cavity case, the flow is turbulent with a Reynolds number equal to 860 000 and compressible with a Mach number of 0.8. Several turbulence models and numerical schemes are used to validate the method. Physical features of the flow are recovered, such as the fundamental frequency of the natural flow as well as acoustic mechanisms, suggesting the validity of the method. The sensitivity gradients are then computed and validated, the error in predicting the eigenvalue variation being found less than 3\%. Control maps using a small steady control device are finally obtained, indicating that the control area should be chosen in the vicinity of the leading edge of the cavity.
\end{abstract}


\end{frontmatter}


\section*{Introduction}

Low frequency unsteady turbulent flows are frequently encountered in engineering applications and generally lead to undesirable features such as structural loads or high level of noise radiation. Predicting and controlling the occurrence of flow unsteadiness is of critical importance in aeronautical applications \cite{GadElHak1998,Collis2004}.

Over the last decades, linear stability analysis appeared to be an adequate tool to characterize laminar flow dynamics. This analysis assumes the existence of a stationary solution $\mathbf{w_{b}}$ of the Navier-Stokes equations upon which a small amplitude unsteady perturbation is added under the form of a normal mode $\mathbf{w}=\hat{\mathbf{w}}e^{\lambda t}$ of spatial structure $\hat{\mathbf{w}}$ and eigenvalue $\lambda$. The evolution equations of the perturbation are given by the linearized Navier-Stokes operator $\mathbf{J}$, the so called Jacobian matrix, and the flow is globally unstable if there exists an exponentially growing mode. Detailed reviews on the characterization of flow dynamics, linking flow unsteadiness to the existence of unstable modes, can be found in \cite{Huerre1990,Godreche1998,Sipp2010}. The role of unstable global modes in flow unsteadiness being more clearly understood, flow control methods targeting the unstable modes were developed in order to manipulate unsteady flows \cite{Chomaz2005,Giannetti2007,Kim2007,Sipp2012}. In particular, prediction of sensitive regions for passive control is of interest as wind tunnel tests and numerical simulations remain expensive. In this spirit, \citet{Marquet2008a} studied the laminar wake behind a two dimensional cylinder for flow parameters above but near the instability threshold ($Re=30-100$). Following previous studies \cite{Hill1992}, they proposed to evaluate the impact on the unstable eigenvalue $\lambda $ of a modification of the baseflow $\mathbf{w_{b}}$ due to the presence of a stationary force $\mathbf{f}$. To this end, they considered the gradient of the unstable eigenvalue with respect to baseflow modifications $\nabla_{\mathbf{w_{b}}}\lambda $, also called the sensitivity gradient to baseflow perturbation, as well as the gradient of the unstable eigenvalue with respect to the introduction of a steady force $\nabla_{\mathbf{f}}\lambda $. Modelling a small cylinder as a  steady force, they predicted the most sensitive regions of the flow to stabilize the unstable global mode and compared their results with the experimental study of  Strykowski \& Sreenivasan \cite{Strykowski1990}. Control maps of both studies overlapped well suggesting that this numerical approach could be a valuable tool in predicting stabilization regions of unsteady flows.

These encouraging results obtained for laminar flow dynamics raised the question of the applicability of such methods for turbulent flows, which are more likely to be encountered in aeronautical applications. Turbulence models remain widely used in this area as the computational cost to solve the Navier-Stokes equations using Direct Numerical Simulation (DNS) drastically increases with the Reynolds number. In the case of turbulent flows for which the scale decoupling assumption holds (see \cite{Rodi1997,Iaccarino2003,Deck2005,Lawson2011}), the dynamics of the large scales of the flow may be captured using unsteady Reynolds-Averaged Navier Stokes (RANS) equations. The impact of the small scales dynamics onto the large ones is accounted for by a turbulence model, which results in additional viscosity (eddy viscosity). Several studies also suggested that linear stability methods based on RANS equations may provide interesting results regarding the underlying mechanism of flow unsteadiness. \citet{Crouch2007,Crouch2009} analysed the buffeting phenomenon for a two dimensional aerofoil. The shock wave starts to oscillate when the angle of attack of the wing and the Mach number reach critical values. They used the one equation turbulence model of Spalart-Allmaras \cite{Spalart1994} and showed that the time integration of the RANS equations reproduced reasonably well the Buffet-onset as well as the frequency of the observed phenomenon. They  linearized the RANS equations and showed that the Buffet onset was linked to the occurrence of an unstable global mode whose frequency matched the expected one. More recently, \citet{Meliga2012} linearized the incompressible RANS equations using the Spalart-Allmaras model to study the dynamics of the wake of a D-shaped cylinder at $R_{e}=13000$. They found that the meanflow (time average of the unsteady flow) was slightly unstable and that the associated global mode was characterized by a frequency corresponding approximately to the one observed experimentally \cite{Sipp2007}. In the spirit of the work of \citet{Marquet2008a}, they analytically derived the sensitivity gradient of the full system of equations. Using a steady force (modelling the presence of a small cylinder) as a means to modify the meanflow, they computed sensitivity maps indicating where the cylinder would efficiently change the frequency of the flow. They compared their results with the experimental study of Parezanovi\'c \& Cadot \cite{Parezanovic2012} who controlled the same flow using a cylinder. Both experimental and numerical sensitivity maps for the frequency change showed reasonably good agreement.

Sensitivity gradients may therefore be a valuable tool for designing open-loop control strategies for both laminar and turbulent flows. Computing sensitivity gradients requires the linearization of the RANS equations, which can either be performed in a continuous framework (the equations are first linearized and then discretized) or in a discrete framework (the equations are first discretized and then linearized). Advantages and drawbacks of both frameworks were early studied in the field of optimal shape design methods \cite{Peter2010} and lead to similar results \cite{Giles2000,Nadarajah2001}. A major advantage of the discrete approach is that the adjoint quantities, which are required to compute the sensitivity gradients, are obtained up to machine precision which is not the case in the continuous case (where they are obtained up to discretization error) \cite{dePando2012}. The discrete framework is also conceptually simpler, since the Jacobian and adjoint matrices are directly defined from the discretized residual $ {\mathcal R} $. In contrast, in the continuous framework, the linearized and adjoint equations need first to be derived, and then discretized, with potentially a different discretization scheme. When discontinuities such as shock waves are present in the flow, a discrete approach based on a shock-capturing method and a conservative scheme automatically yields valid direct and adjoint matrices. In contrast, in a continuous framework, \citet{Giles2001} showed that special care must be taken for the linearized and adjoint equations. If not, \citet{Crouch2007} showed that the shock discontinuities in the baseflow need first to be smoothed for the linear analysis to be valid.

In both the continuous and discrete approaches, analytical derivation of the linearized equations remains a difficult task. Indeed, as noted by \citet{Peter2007}, the governing equations may involve complex equations with turbulence models and complex boundary conditions (characteristic boundary conditions \cite{Poinsot1992}). In a discrete framework, the discretization scheme may also include complex spatial discretization techniques (centered schemes with artificial viscosity \cite{Jameson1981,Lerat2001}, upwing schemes \cite{Roe1981,Vanleer1997} with limiters \cite{Albada1982}).  For example, the analytical derivation of the gradients $\nabla_{\mathbf{w_{b}}}\lambda $ and $\nabla_{\mathbf{f}}\lambda $ in the case of compressible turbulent RANS equations has not yet been achieved although this system of equations is more likely to represent practical aeronautical cases.

We propose in this study a fully discrete framework where the linearized equations are obtained from a finite difference method rather than analytical derivation. We will show how the direct and adjoint global modes as well as the sensitivity gradients can be obtained solely from residual evaluations. In particular, we will show that the sensitivity gradients are linked to the Hessian of the governing equations. Such a procedure avoids complex analytical treatments and can easily handle different systems of equations and different spatial discretization schemes. All the complexity (equations, boundary conditions, spatial discretization scheme) is actually accounted for in the evaluation of the residual equation $\mathcal R$,
which is available in all numerical codes. Hence, we will show how a numerical code can be used in a black box manner to compute global modes, adjoint global modes and sensitivity gradients. The discrete framework based on finite difference evaluations therefore yields a highly flexible strategy which is important since several turbulence models and discretization schemes are generally required to cover a variety of configurations (separation, mixing layers, boundary layers, ...) and regimes (subsonic, transonic, supersonic). Of course, the price to pay is that the various quantities involved in the analysis (Jacobian, adjoint matrix, global modes, adjoint global modes, sensitivity gradients) are computed with some error due to the inherent approximations involved in a finite difference method \cite{dePando2012}. The method will be validated on the compressible RANS equations with two different turbulence models in the case of a well documented deep cavity flow at Mach number $0.80$ and Reynolds number $860 000$, which was experimentally studied by \citet{Forestier2003}. \\

The paper is organised as follows. We first introduce in \S \ref{sec1} the stability theory background and define the sensitivity gradients. In \S  \ref{sec2}, we present the fully discrete approach based on a finite difference technique to compute the direct and adjoint global modes and the gradients. We discuss also various numerical strategies to obtain these quantities, one based on an explicit matrix strategy combined with a direct LU solver (cheap in time, but expensive in memory) and another based on iterative algorithms (cheap in memory, but expensive in time). Technical aspects of the numerical method used to linearize the full system of equations and compute the sensitivity gradients are given in \S  \ref{sec3}. Finally, we will validate this method in the case of a deep cavity flow at high Reynolds number. Case specifications for the validation are presented in \S  \ref{sec4} while linear stability results and sensitivity gradients are discussed in \S  \ref{sec5}.

\newpage

\section{Sensitivity analysis in a discrete framework} \label{sec1}

This section is devoted to the presentation of the linear stability and the sensitivity analyses. We consider generic governing equations of the flow, which encompass in particular the case of a compressible flow whom dynamics is modelled using RANS equations closed with a turbulence model.

\subsection{Linear stability}

After spatial discretization, the governing equations can be recast in the general following conservative form:
\begin{eqnarray}
\dfrac{d\mathbf{w}}{dt} =\mathcal{R}\left(\mathbf{w}\right),  \label{eq1}
\end{eqnarray}
where  $\mathbf{w}\in\mathbb{R}^{N}$ represents the set of conservative variables describing the flow at each spatial location of the mesh and $\cal{R}$ $:$ $\Omega$$\in\mathbb{R}^{\text{N}}\rightarrow\mathbb{R}^{\text{N}}$ is $\cal{C^{\text{2}}}$ over $\Omega$ and represents the discrete residuals. Using finite volume or finite difference methods, the dimension of $\mathbf{w}$ corresponds to the number of cells or nodes in the mesh times the number of variables. Note that all boundary conditions are included in the discrete operator $\cal{R}$.

We assume the existence of a steady solution $\mathbf{w_{b}}\in\mathbb{R}^{N}$ to this system referred to as the baseflow and defined by the discrete equation:
\begin{equation} \label{baseflow}
\cal{R}\left(\mathbf{w_{b}}\right)=\mathbf{0}.
\end{equation}
In the case of governing equations involving a turbulence model, it is worth mentioning that such a baseflow
takes into account the Reynolds stresses involved in the turbulence model, but not those related
to possible low-frequency (and large-scale) perturbations, which are accounted for by the time-integration
in Eq. (\ref{eq1}). In so far, the above defined baseflow is not strictly speaking a meanflow (even though
it incorporates some meanflow effects due to high-frequency turbulence) and
may therefore be considered as a valid candidate for a stability analysis.

The stability of the baseflow is probed by analysing the evolution of a small amplitude perturbation $\epsilon\mathbf{w}'$
superimposed on the baseflow: $\mathbf{w}=\mathbf{w_{b}}+\epsilon\mathbf{w}'$, with $ \epsilon \ll 1 $. Note that in the case of governing equations involving
a turbulence model, the perturbation also involves variations of the turbulent quantities.
The equation governing the perturbation is given by the linearization to the first order of the discretized
equations in (\ref{eq1}):
\begin{equation}
\dfrac{d\mathbf{w}'}{dt}={\mathbf J}\mathbf{w}'. \label{eq2}
\end{equation}
The Jacobian operator ${\mathbf J}\in\mathbb{R}^{\text{N}\times\text{N}}$ corresponds to the linearization of the discrete  Navier-Stokes operator $\mathcal{R}$ around the baseflow $\mathbf{w_{b}}$:
\begin{equation} \label{jac}
\mathbf{J}_{{ij}}=\left.\dfrac{\partial{\mathcal R}_{{i}}}{\partial {\mathbf w}_{{j}}}\right|_{\mathbf{w}=\mathbf{w_{b}}},
\end{equation}
where $ {\mathcal R}_{{i}} $ designates the $i^{\text{th}} $ component of the residual, which is a priori
a function of all unknowns ${\mathbf w}_{{j}} $ in the mesh.
If we use finite volume or finite difference methods, then the spatial discretization stencil is compact and
the $i^{\text{th}} $ component of the residual only depends on few neighbouring unknowns.
Hence, $\mathbf{J}$ is a sparse matrix in such cases.

We consider perturbations under the form of normal modes $\mathbf{w}'=\hat{\mathbf{w}}e^{\lambda t}$, where $\lambda = \sigma + i\omega$ describes its temporal behaviour --- $\sigma$ is the amplification rate and $\omega $ the frequency --- and $\hat{\mathbf{w}}\in\mathbb{C}^{N}$ its spatial structure. Then  Eq. (\ref{eq2}) may be recast into the following eigenvalue problem:
\begin{equation} \label{linear1}
{\mathbf J}\hat{\mathbf{w}}=\lambda \hat{\mathbf{w}}.
\end{equation}
If at least one of the eigenvalues $ \lambda $ exhibits a positive growth rate $\sigma$, then the baseflow $ \mathbf{w_b} $ is unstable.\\

\textbf{Remark:} We assumed the residual operator $\cal{R}$ to be $\cal{C^{\text{2}}}$$\left(\Omega\right)$ for the sensitivity gradients to be defined. Strictly speaking, the stability analysis only requires the considered system of equations to be differentiable, that is $\cal{R}\in\cal{C^{\text{1}}}$$\left(\Omega\right)$.

\subsection{Sensitivity study}

Let us consider a particular eigenmode  $\left(\lambda{,}\hat{\mathbf{w}}\right)$. Following previous studies \cite{Marquet2008a,Hill1992,Bottaro2003}, this eigenmode may be considered as a function of the baseflow $ \mathbf{w_{b}} $, since the Jacobian matrix has been obtained by linearization of the governing equations near the baseflow. Hence, a small baseflow perturbation $\boldsymbol{\delta} \mathbf{w_{b}}$ generates a small variation of the eigenvalue $\delta \lambda$, which can be written as:
\begin{eqnarray} \label{sensitivity1}
\delta\lambda=\left<\boldsymbol{\nabla}_{\mathbf{w_{b}}}\lambda{,}\;\boldsymbol{\delta}\mathbf{w_{b}}\right> .
\end{eqnarray}
This expression defines the gradient $\boldsymbol{\nabla}_{\mathbf{w_{b}}}\lambda\in\mathbb{C}^{\text{N}}$, called the sensitivity of the eigenvalue to baseflow modifications. It is a complex vector field, the real and imaginary parts respectively dealing with the sensitivity of the amplification rate and the frequency.
Note that in the case of governing equations including a turbulence model, one may analyse the sensitivity of the
global mode to variations of turbulent scales of the baseflow.
In Eq. (\ref{sensitivity1}), the discrete inner product $\left< \cdot \right>$ refers to the Euclidian inner-product in $\mathbb{C}^{\text{N}}$:
\begin{equation} \label{pscanonical}
\left< \mathbf{u}, \mathbf{v} \right>=\mathbf{u}^*\mathbf{v},
\end{equation}
where $ ^* $ denotes conjugate transpose. The associated norm $\left\|\mathbf{u}\right\|=\sqrt{\left< \mathbf{u}, \mathbf{u} \right>}$ will be used in the following.

We now derive an explicit expression of $\boldsymbol{\nabla}_{\mathbf{w_{b}}}\lambda$. Note again that this has been done up to now in a continuous framework, while the goal of the present paper is to introduce the discrete one.
First, let us recall that an arbitrary variation of the Jacobian $\boldsymbol{\delta}{\mathbf J}$ induces the following variation of the eigenvalue $\delta\lambda$ \cite{Sipp2010}:
\begin{equation} \label{firstDeriv0}
\delta\lambda=\left<\tilde{\mathbf{w}}{,}\;\boldsymbol{\delta}{\mathbf J}\hat{\mathbf{w}}\right>,
\end{equation}
where $\tilde{\mathbf{w}}\in\mathbb{C}^{N}$ corresponds to the adjoint global mode, solution of the following eigenproblem:
\begin{equation}  \label{adjoint}
{\mathbf J}^*\tilde{\mathbf{w}}=\lambda^{*}\tilde{\mathbf{w}} \hspace{0.5cm} \text{with} \hspace{0.5cm} \left<\tilde{\mathbf{w}}{,}\hat{\mathbf{w}}\right>=\text{1.}
\end{equation}

If $\boldsymbol{\delta}{\mathbf J}$ corresponds to a variation of the Jacobian induced by a variation of the baseflow $\boldsymbol{\delta} \mathbf{w_{b}}$, then:
\begin{equation}
 \boldsymbol{\delta}{\mathbf J}\hat{\mathbf{w}}=\left.\dfrac{\partial ({\mathbf J} \hat{\mathbf{w}})}{\partial \mathbf{w}}\right|_{ \mathbf{w}=\mathbf{w_{b}}}\boldsymbol{\delta} \mathbf{w_{b}},
\end{equation}
where the global mode $ \hat{\mathbf{w}} $ is assumed to be frozen.
This expression may be written in a different manner using the Hessian $\mathbf{H}$ of $\mathcal{R}$:
\begin{equation} \label{dJ}
 \boldsymbol{\delta}{\mathbf J}\hat{\mathbf{w}}=\mathbf{H}(\hat{\mathbf{w}},\boldsymbol{\delta}\mathbf{w_{b}}).
\end{equation}
Here $ \mathbf{H}\left(\mathbf{u},\mathbf{v}\right)\in\mathbb{C}^{\text{N}}$ designates
the vector $\mathbf{z}$ such that ${\mathbf z}_{{i}}=\sum_{j,k}{\mathbf{H}_{ijk}\mathbf{u}_{j}\mathbf{v}_{k}}$, with:
\begin{equation}  \label{Hessian}
\mathbf{H}_{{ijk}}=\left. \dfrac{\partial^{{2}}\mathcal{R}_{{i}}}{\partial\mathbf{w}_{j}\partial\mathbf{w}_{k}}\right|_{\mathbf{w}=\mathbf{w_{b}}}.
\end{equation}
Similarly to the discussion for the Jacobian $ \mathbf{J} $,
if compact differential stencils are used, then for each component $ i $
only few values of $\mathbf{H}_{{ijk}}$ are non-zero.

Let us introduce the matrix $\mathbf{H}^{\prime}\in\mathbb{C}^{\text{N}\times\text{N}}$ such that $\mathbf{H}^{\prime}\boldsymbol{\delta} \mathbf{w_{b}}=\mathbf{H}(\hat{\mathbf{w}},\boldsymbol{\delta} \mathbf{w_{b}})$ for all $\boldsymbol{\delta} \mathbf{w_{b}}$. Hence:
\begin{equation} \label{Hprime}
 {\mathbf H}^{\prime}_{ik}=\sum_j {\mathbf H}_{ijk}\hat{\mathbf{w}}_j=\left<\mathbf{e_i}\text{,}\mathbf{H}(\hat{\mathbf{w}},{\mathbf e}_k)\right>.
\end{equation}
Here, ${\mathbf e}_i$ denotes the unit vector on the $i^{th}$ component of the canonical basis of $\mathbb{R}^{\text{N}}$. Equation (\ref{dJ}) may then be rewritten as:
\begin{equation} \label{dJv2}
 \boldsymbol{\delta}{\mathbf J}\hat{\mathbf{w}}={\mathbf H}' \boldsymbol{\delta} \mathbf{w_{b}}.
\end{equation}
Introducing Eq. (\ref{dJv2}) into (\ref{firstDeriv0}), we have:
\begin{equation} \label{sensitivity1b}
\delta\lambda =\left<\tilde{\mathbf{w}}{,}{\mathbf H}'\boldsymbol{\delta}\mathbf{w_{b}} \right> =\left<{\mathbf H}^{\prime*}\tilde{\mathbf{w}}{,}\boldsymbol{\delta}\mathbf{w_{b}} \right> .
\end{equation}
If we identify this expression with Eq. (\ref{sensitivity1}), we obtain the following expression of the gradient:
\begin{equation}  \label{gradient1}
\boldsymbol{\nabla}_{\mathbf{w_{b}}}\lambda=\mathbf{H}^{\prime*}\tilde{\mathbf{w}}.
\end{equation}

In the view of open-loop control that aims at stabilizing the unstable global modes,
we will consider control devices that act by adding volumic source terms
to the Navier-Stokes equations. For example, any object in the flow
may be represented as a force, while heating or cooling is
a source term in the energy equation. If a turbulence model is considered in the
governing equations, then control devices that locally modify the turbulent scales of the flow may also
be considered.
In the following, we consider the impact of a small amplitude source term $\boldsymbol{\delta}\mathbf{f}\in\mathbb{R}^{N}$, which modifies the baseflow such that $\mathcal{R}\left(\mathbf{w_{b}}+\boldsymbol{\delta}\mathbf{w_{b}}\right)+\boldsymbol{\delta}\mathbf{f}=\mathbf{0}$. Linearising this expression about $\mathbf{w_{b}}$, we obtain the baseflow modifications due to the small amplitude source term: $\boldsymbol{\delta}\mathbf{w_{b}}=-{\mathbf J}^{{-1}}\boldsymbol{\delta} \mathbf{f}$.
Rewriting equation (\ref{sensitivity1b}), we obtain:
\begin{equation}
\delta\lambda =\left<{\mathbf H}^{\prime*}\tilde{{\mathbf w}}{,}-{\mathbf J}^{{-1}}\boldsymbol{\delta} \mathbf{f} \right> =
\left<-{\mathbf J}^{*-1}{\mathbf H}^{\prime*}\tilde{\mathbf{w}}{,}\boldsymbol{\delta} \mathbf{f} \right> .
\end{equation}
The sensitivity of the eigenvalue to the introduction of a source term $\boldsymbol{\nabla}_{\mathbf{f}}\lambda\in\mathbb{C}^{N}$, which links the eigenvalue variation $\delta\lambda$ to the steady source term $\boldsymbol{\delta}\mathbf{f}$, is thus given by:
\begin{eqnarray} \label{gradient2}
\delta\lambda=\left<\boldsymbol{\nabla}_{\mathbf{f}}\lambda{,}\boldsymbol{\delta}\mathbf{f}\right>\hspace{0.5cm} \text{with} \hspace{0.5cm} \boldsymbol{\nabla}_{\mathbf{f}}\lambda=-{\mathbf J}^{*-1}\boldsymbol{\nabla}_{\mathbf{\mathbf{w_{b}}}}\lambda.
\end{eqnarray}
The impact of a small amplitude steady source term on the flow spectrum can therefore be predicted a priori and control maps can be obtained beforehand.

We used up to now the canonical inner product (\ref{pscanonical}).
Yet, to give physical meaning to the gradient so as to allow comparisons of results (if available) with those obtained from a continuous approach, it may be useful to choose another inner-product, based on a positive definite hermitian matrix $ \mathbf Q$ such that:
\begin{eqnarray} \label{ps}
\left.\left<{\mathbf u}{,}{\mathbf v}\right>\right|_{\mathbf Q}={\mathbf u}^*{\mathbf Q}{\mathbf v},
\end{eqnarray}
with the corresponding norm $\left\|\mathbf{u}\right\|_{\mathbf{Q}}=\sqrt{\left<{\mathbf u}\text{,}{\mathbf u}\right> |_{\mathbf{Q}}}$.
Based on this new inner-product, the sensitivities may be defined as follows
\begin{eqnarray}
\delta\lambda&=&\left.<\left.\boldsymbol{\nabla}_{\mathbf{w_{b}}}\lambda\right|_{\mathbf Q}{,}\;\boldsymbol{\delta}\mathbf{w_{b}}>\right|_{\mathbf Q} \\
             &=&\left.<\left.\boldsymbol{\nabla}_{\mathbf{f}}\lambda\right|_{\mathbf Q}{,}\;\boldsymbol{\delta}\mathbf{f}>\right|_{\mathbf Q},
\end{eqnarray}
and one straightforwardly obtains:
\begin{eqnarray} \label{gradient1b}
\left.\boldsymbol{\nabla}_{\mathbf{w_{b}}}\lambda\right|_{\mathbf Q} &=& {\mathbf Q}^{-1} \boldsymbol{\nabla}_{\mathbf{w_{b}}}\lambda \\ \label{gradient2b}
\left.\boldsymbol{\nabla}_{\mathbf{f}}\lambda\right|_{\mathbf Q} &=& {\mathbf Q}^{-1} \boldsymbol{\nabla}_{\mathbf{f}}\lambda .
\end{eqnarray}
For sake of completeness, the adjoint global mode associated to this new inner-product is:
\begin{eqnarray} \label{adjointQ}
\left. \tilde{\mathbf{w}} \right|_{\mathbf{Q}}={\mathbf Q}^{-1} \tilde{\mathbf{w}}.
\end{eqnarray}

\noindent To sump up, in order to compute the sensitivity gradients, we need to compute:
\begin{enumerate}
 \item unstable direct global modes $\hat{\mathbf{w}}$ based on the discrete Jacobian $ \mathbf{J} $ (Eq. (\ref{linear1}));
 \item unstable adjoint global modes $\tilde{\mathbf{w}}$ based on the discrete adjoint Jacobian $ \mathbf{J}^*$ (Eq. (\ref{adjoint}));
 \item $\mathbf{H}^{\prime*} \tilde{\mathbf{w}}$ (see Eq. (\ref{Hprime}) for the definition of $ \mathbf{H}' $) to obtain the sensitivity of the global mode to baseflow modifications $ \boldsymbol{\nabla}_{\mathbf{w_{b}}}\lambda $ (Eq. (\ref{gradient1}));
 \item $-{\mathbf J}^{*-1} \boldsymbol{\nabla}_{\mathbf{w_{b}}}\lambda $ to obtain the sensitivity of the global mode to the introduction of a steady source term $ \boldsymbol{\nabla}_{\mathbf{f}}\lambda $ (Eq. (\ref{gradient2})).
 \item $ \left.\boldsymbol{\nabla}_{\mathbf{w_{b}}}\lambda\right|_{\mathbf Q}={\mathbf Q}^{-1}\boldsymbol{\nabla}_{\mathbf{w_{b}}}\lambda $ (Eq. (\ref{gradient1b})) and $ \left.\boldsymbol{\nabla}_{\mathbf{f}}\lambda\right|_{\mathbf Q}={\mathbf Q}^{-1}\boldsymbol{\nabla}_{\mathbf{f}}\lambda $
 (Eq. (\ref{gradient2b})) to obtain sensitivities with a physically relevant inner-product (\ref{ps}).
\end{enumerate}

\section{Numerical strategy} \label{sec2}

The procedure to compute the sensitivity gradients presented in \S \ref{sec1} relies on the knowledge
of the first (the Jacobian $\mathbf{J}$) and second (the Hessian $\mathbf{H} $) derivatives
of the discrete operator ${\mathcal R}(\mathbf{w})$. As mentioned in the Introduction, we follow in this article a strategy based on a finite difference method to obtain both $\mathbf{J}\mathbf{u}$ and $\mathbf{H}({\mathbf u},{\mathbf v}) $ with $\mathbf u$ and $\mathbf v$ arbitrary vectors. More precisely, we want to evaluate these matrices by repeated evaluations of the residual function. The code may then be used in a black box manner: assuming that the code generates
a valid discrete residual $ {\mathcal R}({\mathbf u}) $,
one may obtain approximations of $\mathbf{J}\mathbf{u}$ and $\mathbf{H}({\mathbf u},{\mathbf v}) $ with the following
first order approximations:
\begin{eqnarray} \label{discjaco}
\mathbf{J}\mathbf{u}&=&\dfrac{\text{1}}{\epsilon}\left[{\mathcal R}\left(\mathbf{w_{b}}+\epsilon\mathbf{u}\right)-{\mathcal R}\left(\mathbf{w_{b}}\right)\right], \\ \label{dischessian}
{\mathbf H}\left(\mathbf{u},\mathbf{v}\right)&=& \dfrac{\text{1}}{\epsilon_1\epsilon_2}\text{[}{\mathcal R}\left(\mathbf{w_{b}}+\epsilon_1\mathbf{u}+\epsilon_2\mathbf{v}\right)
- {\mathcal R}\left(\mathbf{w_{b}}+\epsilon_1\mathbf{u}\right) \\
&\quad{}& \quad{} -{\mathcal R}\left(\mathbf{w_{b}}+\epsilon_2\mathbf{v}\right)+{\mathcal R}\left(\mathbf{w_{b}}\right)\text{]}, \notag
\end{eqnarray}
where $ \epsilon $, $\epsilon_1$ and $ \epsilon_2 $ are small constants.
The choice of these constants will be further detailed in \S \ref{sec3}.
In the context of global stability analyses, finite difference methods have already been
used to approximate the discrete Jacobian \cite{dePando2012,mack2010}.
Here, we suggest that these methods may also be useful to compute the sensitivity gradients introduced
in \S \ref{sec1}.

To validate this idea, we have chosen an "explicit matrix" approach combined with a direct sparse LU solver to perform matrix inversions, which is relevant for small-scale-problems of the order of $ 10^6-10^7 $ degrees of freedom for ${\mathbf w}$. The advantage of this strategy is that it yields fast and accurate results.
The "explicit matrix" strategy consists in computing and storing all non-zero values of the various
matrices involved in \S \ref{sec1}.
Due to the large size of the meshes this is possible only if these matrices are sparse.
The Jacobian structure is intrinsically linked to the stencil width used to discretize $\mathcal{R}$, which we assume to be compact, ensuring the sparse nature of $\mathbf{J}$.
Moreover, a similar result holds for matrix $ \mathbf{H}' $ (see next section for details). Explicit knowledge
of these matrices induces that we also have direct access
to $ {\mathbf J}^* $ and
$ {\mathbf H}^{\prime *} $ involved in steps 2, 3 and 4
of the procedure summarized at the end of \S \ref{sec1}.
Both eigenvalue problems in Eqs. (\ref{linear1}) and (\ref{adjoint}) may be solved using Krylov methods with a shift-invert strategy (open source library ARPACK \cite{lehoucq1998}), so as to focus on the least-damped eigenvalues.
Matrix inversions involved in these eigenproblems and in step 4
of the procedure outlined at the end of \S \ref{sec1}
are carried out in the following with a direct sparse LU solver for distributed memory machines (MUMPS see http://graal.ens-lyon.fr/MUMPS/, or SuperLU-dist see http://acts.nersc.gov/superlu/).
The inverses are obtained extremely fast but the drawback is
the very high requirements in terms of memory (typically
around 50 times the memory of the matrix to be inverted).
In order to avoid this overshoot in memory, one could
use, instead of the direct LU solvers, iterative algorithms such as BICGSTAB with an incomplete LU preconditioner \cite{mack2010}. This would however result in a strong increase in computational time.

For problems with a larger number of degrees of freedom, typically 3D problems, one has to resort
to "on the fly" approaches, where the matrix is never stored explicitly. The "on the fly" strategy has been introduced in the context of global stability analyses by Mamun  et al. \cite{mamun1995},  \citet{bagheri2009aiaaj} and Mack et al. \cite{mack2010}.
The objective here is to avoid forming any matrix explicitly
in order to save memory. This requires specific
algorithms that are solely based on the action
of the matrices on a vector.
\citet{dePando2012} have shown in the context of laminar compressible
flows how to efficiently compute $ \mathbf{J}\mathbf{u} $
and $ \mathbf{J}^*\mathbf{u} $ by using finite differences with
an existing direct numerical simulation code.
Also, they showed that time-integration of Eq. (\ref{eq2}) combined with a Krylov-Schur method and a Harmonic extraction technique
effectively recovered the least-damped direct and adjoint global modes.

However, these previous studies  using "on the fly" strategy were not concerned with the computation of the sensitivity gradients. We shall remark here that in Step 3, which is devoted to the computation of the sensitivity
to baseflow modifications, the evaluation of ${\mathbf z}={\mathbf H}^{\prime*} \tilde{\mathbf{w}} $ can in principle also be performed "on the fly":
\begin{eqnarray}
{\mathbf z}_i&=&\sum_j \overline{{\mathbf H}^{\prime}_{ji}}\tilde{\mathbf{w}}_j =\sum_j \overline{\left< {\mathbf e}_j \text{,} \mathbf{H}(\hat{\mathbf{w}},{\mathbf e}_i) \right>}\tilde{\mathbf{w}}_j \\
&=&\overline{\left< \tilde{\mathbf{w}}\text{,} \mathbf{H}(\hat{\mathbf{w}},{\mathbf e}_i) \right>}.
\end{eqnarray}
where $ \mathbf{H}(\hat{\mathbf{w}},{\mathbf e}_i) $ can be approximated from Eq. (\ref{dischessian}).
This evaluation may be computationally intensive since one Hessian evaluation $\mathbf{H}(\hat{\mathbf{w}},{\mathbf e}_i)$ has to be performed by degree of freedom
so that some optimization may be useful. Nevertheless, since this evaluation
is only done once per considered eigenmode, it is less critical than the
evaluations of ${\mathbf J}{\mathbf u} $ and ${\mathbf J}^*{\mathbf u} $, required for the time-integration in
the eigenproblems.

Note that the inversions $ {\mathbf Q}^{-1}{\mathbf u} $ involved in step 5 may easily be carried out with a cheap conjugate gradient algorithm with diagonal preconditioning.

\section{Efficient evaluation of matrices with explicit storage}
\label{sec3}

The procedure used to efficiently compute the matrices $\mathbf{J}$ and $\mathbf{H}^{\prime}$ by taking advantage of their structure dependence to the discretization stencil is first detailed. The choice of the linearization parameters is then discussed. A more intrusive approach suited for codes containing an existing linearization of the RANS equations such as shape optimization codes will finally be investigated. We consider in the following a case of dimension $d$ solved using finite volume or finite difference methods with a discretization scheme using an $n_{s}$ points stencil in each direction. We assume the governing equations (\ref{eq1}) to be discretized on a mesh of size $N_{m}=I_{m}\times J_{m}\times K_{m}$ for a system of $n_{c}$ conservative variables. As will be further detailed, the Jacobian is a square matrix of size $N\times N$ where $N=n_{c}\times N_{m}$, with a total number of non zero elements $n_{e}$.

\subsection{Example case}

As an example case, we consider the following $d=1$ dimensional model with $n_{c}=2$ two conservative variables discretized on an $n_{s}=2$ points stencil, the discretization step $\Delta x$ being taken uniform and equal to $1$ for simplicity :
\begin{eqnarray}
\mathcal{R}\left(\mathbf{w }\right)=
\mathcal{R}
\begin{pmatrix}
\mathbf{a} \\
\mathbf{b}
\end{pmatrix}
=
\begin{pmatrix}
\mathbf{b}\partial_{x}\mathbf{a}\\
\mathbf{a} \partial_{x}\mathbf{b}
\end{pmatrix}
\hspace{1cm}
\mathcal{R_{\text{i}}}=
\begin{pmatrix}
b_{i}\left[a_{i+1}-a_{i}\right] \\
a_{i}\left[b_{i+1}-b_{i}\right]
\end{pmatrix}.
\end{eqnarray}

Linearizing the discrete equations, we obtain the product $\mathbf{J}\mathbf{u}$ in the stencil $\left( i,i+1,i+2\right)$:

\begin{equation} \label{discexample}
\underbracket[1pt][6pt]{
\begin{pmatrix}
 -b_{i} & a_{i+1}-a_{i} & b_{i}& 0& 0& 0 \\
 b_{i+1}-b_{i} & -a_{i}& 0& a_{i}& 0& 0 \\
 0& 0 & -b_{i+1} & a_{i+2}-a_{i+1} & b_{i+1} & 0   \\
 0& 0 & b_{i+2}-b_{i+1}& -a_{i+1}& 0 & a_{i+1}
\end{pmatrix}
}_{\mathbf{J}}
\underbracket[1pt][1pt]{
\begin{pmatrix}
da_{i} \\
db_{i} \\
da_{i+1} \\
db_{i+1} \\
da_{i+2} \\
db_{i+2}
\end{pmatrix}
}_{\mathbf{u}}
\end{equation}

We foresee that all the Jacobian coefficients can be obtained independently from this matrix vector product using the following set of vectors $\mathbf{u}$ : \\
\begin{equation}
\mathbf{u}=
\begin{pmatrix}
 \vdots \\ da_{i-1} \\ db_{i-1} \\ da_{i}  \\ db_{i} \\  da_{i+1} \\ db_{i+1} \\ da_{i+2} \\ db_{i+2} \\ \vdots
\end{pmatrix}
=\underbrace{\begin{pmatrix}
 \vdots \\ 0 \\ 0 \\ 1   \\ 0 \\  0 \\ 0 \\ 1 \\ 0 \\ \vdots
\end{pmatrix}
=\begin{pmatrix}
 \vdots \\ 1 \\ 0 \\ 0   \\ 0 \\  1 \\ 0 \\ 0 \\ 0 \\ \vdots
\end{pmatrix}
}_{\text{da}}
=\underbrace{\begin{pmatrix}
 \vdots \\ 0 \\ 0 \\ 0   \\ 1 \\  0 \\ 0 \\ 0 \\ 1 \\ \vdots
\end{pmatrix}
=\begin{pmatrix}
 \vdots \\ 0 \\ 1 \\ 0   \\ 0 \\  0 \\ 1 \\ 0 \\ 0 \\ \vdots
\end{pmatrix}
}_{\text{db}}
\end{equation}

This set corresponds to perturbation vectors $\mathbf{e_{i}}$ taken every $n_{s}=2$ points for each variable $\mathbf{a}$ and $\mathbf{b}$ separately. The non zero indices in the perturbation vectors are shifted every $n_{c}\times n_{s}=4$ points to ensure that we only compute one contributing term ($da_{i}$, $da_{i+1}$, $db_{i}$ or $db_{i+1}$ for example) for each matrix vector product. The Jacobian can thus be obtained using $n_{s}\times n_{c}=4$ residual evaluations. Each line of the Jacobian contains $n_{c}\times n_{s}$ non zero coefficients, we thus have $n_{e}\approx n_{c}\times n_{s}\times N =8I_{m}$.

\subsection{General procedure}

The Jacobian is computed according to Eq. (\ref{discjaco}) by evaluation of the discrete residuals at each point. Using an $n_{s}$ points stencil, the discrete residual at point $\left(i,j,k\right)$ for the $v^{\text{th}}$ variable $\cal{R_{\text{ijk}}^{\text{v}}}=\cal{R}\left(\mathbf{W}_{\text{lmn}}\right)$ is only a function of the $\left(\text{l,m,n}\right)$ points linked to $\left(i,j,k\right)$ by the discretization stencil that is at most $n_{s}^{d}$ points ($d=1\text{,}2\text{,}3$ if we consider respectively a one, two or three dimensional case). As an example, a two dimensional case with $n_{s}=5$ (see \S \ref{sec4}) is depicted in Fig.\ref{fig1}(a) where the dependency of the residual $\cal{R_{\text{ij}}^{\text{v}}}$ towards the stencil is plotted. We foresee from this example that the total number of points $n_{p}$ which contribute to the residual evaluation at one point may differ from the maximum value $n_{s}^{d}$, that is $n_{p}\le n_{s}^{d}$ (in the Figure $n_{p}=13$ while $n_{s}^{2}=25$).

The Jacobian coefficients can be interpreted as the contribution of the $\left(l,m,n\right)$ point to the linearization around the baseflow of the discretized equations at the point $\left( i,j,k\right)$. Linearizing the equations at $\left( i,j,k\right)$ for a given variable, we obtain $n_{c}$ coefficients for each of the $n_{p}$ contributing points $\left( l,m,n\right)$. Therefore the  total number of non zero elements in the Jacobian scales as  $n_{e}\approx n_{p}n_{c}N =n_{p}n_{c}^{2}N_{m}$. Note that $n_{e}$ corresponds to the maximum number of non zero elements in the matrix and may overpredict the actual number.  The sparsity coefficient of the matrix $S=1-n_{e}/ N^{2}\approx 1-n_{p} /N_{m}$ is reduced when the stencil width of the system is increased.

Perturbing the baseflow with a vector $\mathbf{e_{lmn}}$ equal to $1$ for a given conservative variable at a point $\left(l,m,n\right)$ and $0$ elsewhere, Eq. (\ref{discjaco}) becomes:
\begin{equation} \label{discresidual}
\mathbf{J}\mathbf{e_{lmn}}=\dfrac{\text{1}}{\epsilon}\left[\cal{R}\left(\mathbf{w_{b}}+\epsilon\mathbf{e_{lmn}}\right)-\cal{R}\left(\mathbf{w_{b}}\right)\right].
\end{equation}
Due to the stencil dependency, the perturbation only impacts the discrete residuals at the $n_{p}$ points around $\left(l,m,n\right)$ in their evaluation. Therefore, the right hand side of the previous equation yields $n_{p}n_{c}$ non zero coefficients of $\mathbf{J}$. These terms correspond to the contribution of $\left(l,m,n\right)$ to the linearization of the equations at these $n_{p}$ points. Therefore the complete linearization of the discrete equations at a point $\left(i,j,k\right)$ can be obtained by perturbing individually all the $n_{p}$ points that intervene in the residual evaluation at $\left(i,j,k\right)$ for each conservative variable. The Jacobian coefficients can thus be obtained independently using Eq. (\ref{discresidual}) by defining a set of perturbation vectors  $\left(\mathbf{e_{p}}\right)$ for each conservative variable and every $n_{s}$ points in each direction. The matrix is obtained by performing  $n_{res}=n_{c}n_{s}^{d}$ residual evaluations (or matrix vector products) and then assembling it explicitly. We shall note here that the residual evaluations for each perturbation vector $\mathbf{e_{p}}$ are independent from one to another: the computational time of this procedure can be greatly lowered using parallel computation. \\

As detailed in \S \ref{sec1}, the computation of the sensitivity gradients mainly requires the computation of the matrix $\mathbf{H}^{\prime}$. As the structure of $\mathbf{H}^{\prime}$ depends on the discretization stencil similarly to that of $\mathbf{J}$, a similar perturbation method may be used to compute it. In particular, using Eq. (\ref{Hprime}) we have:
\begin{eqnarray} \label{fulldiscHessian}
\mathbf{H}^{\prime}\mathbf{e_{p}}&=& \mathbf{H}\left(\hat{\mathbf{w}}\text{,}\mathbf{e_{p}}\right) \\ \notag
&=&\dfrac{\text{1}}{\epsilon_{\text{1}}\epsilon_{\text{2}}}[\cal{R}\left(\mathbf{w_{b}}+\epsilon_{\text{1}}\hat{\mathbf{w}}+\epsilon_{\text{2}}\mathbf{e_{p}}\right)-\cal{R}\left(\mathbf{w_{b}}+\epsilon_{\text{1}}\hat{\mathbf{w}}\right) \\ \notag
&\quad& \quad \quad -\cal{R}\left(\mathbf{w_{b}}+\epsilon_{\text{2}}\mathbf{e_{p}}\right)+\cal{R}\left(\mathbf{w_{b}}\right)],  \notag
\end{eqnarray}
where $\left(\mathbf{e_{p}}\right)$ corresponds to the set of perturbation vectors previously defined. The size of $\mathbf{H}^{\prime}$ and its number of non zero elements are thus equal to the Jacobian ones. The computational cost of explicitly forming $\mathbf{H}^{\prime}$ is four times the Jacobian one as two complex residual evaluations have to be performed for each $\mathbf{e_{p}}$ in Eq. (\ref{fulldiscHessian}). In a code where only real structures are available, all the above mentioned evaluations shall be done separately for both real and imaginary parts of the eigenmode $\hat{\mathbf{w}}$. Indeed, as $\boldsymbol{\delta}{\mathbf J}$ and $\boldsymbol{\delta}\mathbf{w_{b}}$ are real quantities in Eq. (\ref{dJv2}), both real and imaginary parts of Eq. (\ref{dJv2}) can be computed separately.

We previously introduced first order linearization formulas for explanation purpose. In practice, second order formulas are used for the computation of both $\mathbf{J}$ and $\mathbf{H}^{\prime}$:
\begin{eqnarray} \label{discresidual2}
\mathbf{J}\mathbf{e_{p}}&=&\dfrac{\text{1}}{2\epsilon}\left[ \cal{R}\left(\mathbf{w_{b}}+\epsilon\mathbf{e_{p}}\right)-\cal{R}\left(\mathbf{w_{b}}-\epsilon\mathbf{e_{p}}\right)\right], \\ \label{fulldiscHessian2}
\mathbf{H}^{\prime}\mathbf{e_{p}}&=&\dfrac{\text{1}}{\text{4}\epsilon_{\text{1}}\epsilon_{\text{2}}}[\cal{R}\left(\mathbf{w_{b}}+\epsilon_{\text{1}}\hat{\mathbf{w}}+\epsilon_{\text{2}}\mathbf{e_{p}}\right)-\cal{R}\left(\mathbf{w_{b}}+\epsilon_{\text{1}}\hat{\mathbf{w}}-\epsilon_{\text{2}}\mathbf{e_{p}}\right) \\ \notag
&\qquad{}&\quad{}-\cal{R}\left(\mathbf{w_{b}}-\epsilon_{\text{1}}\hat{\mathbf{w}}+\epsilon_{\text{2}}\mathbf{e_{p}}\right)+\cal{R}\left(\mathbf{w_{b}}-\epsilon_{\text{1}}\hat{\mathbf{w}}-\epsilon_{\text{2}}\mathbf{e_{p}}\right)].
\end{eqnarray}

For both matrix computations, the second order precision procedure is twice more costly than the first order one as twice more residual evaluations have to be performed for each $\mathbf{e_{p}}$.

\subsection{Adequate choice of linearization parameters}

The linearization parameters $\epsilon\text{,}\epsilon_{1}\text{,}\epsilon_{2}$ in Eqs. (\ref{discresidual}-\ref{fulldiscHessian2}) should not be too small to avoid round-off errors
and not too large for the approximations to remain accurate. This issue and optimal choices of $ \epsilon $ have been discussed in detail by Knoll et al. \cite{knoll2004} in the context of Jacobian free methods.

Here, we compute each coefficient of the Jacobian individually, so that we actually linearize a scalar equation. The linearizaton parameter can thus be taken as mentioned in \cite{knoll2004} : $\epsilon=\epsilon_{m}\left(\left|w\right|+1\right)$, with $w$ the local baseflow value of the considered variable. Noting $M_{p}$ the machine precision (64 bit machines), \citet{An2011} showed that the $\epsilon_{m}$ which minimized the error should be taken equal to $\epsilon_{m}=\sqrt{M_{p}}\approx 10^{-8}$ for the first order approximation, and equal to $\epsilon_{m}=\sqrt[3]{M_{p}/2}\approx 5.10^{-6}$ for second order ones. Note that when performing second order precision computations, as some conservative variables should remain positive by definition, the imposed perturbation must remain smaller than the baseflow local value. When the previous choice of $\epsilon$ does not satisfy this criterion, we imposed $\epsilon$ to be 10 times smaller than the local baseflow value $\left|w\right|$.

For the computation of $\mathbf{H}^{\prime}$ which is a second order derivative, $\epsilon_{1}$ is taken (similarly to Jacobian free methods \cite{mack2010}) such that the unstable mode $\epsilon_{1}\hat{\mathbf{w}}$ can be considered as small compared to the baseflow $\epsilon_{1}\left\|\hat{\mathbf{w}}\right\|=\sqrt{M_{p}}\left\|\mathbf{w_{b}}\right\|$. This choice of global $\epsilon_{1}$ ensures that the matrix can be computed in $n_{c}n_{s}^{d}$ residual evaluations. In the case of a local $\epsilon_{1}$ (where the value of $\epsilon_{1}$ may differ from one point to another), each local contribution $\mathbf{H_{ijk}}\hat{\mathbf{w_j}}$ in Eq. (\ref{Hprime}) should be computed independently. This can be done by defining a set of vectors $\left(\mathbf{e_{p}}^{\prime}\right)$ with the same structure as the previously defined set $\left(\mathbf{e_{p}}\right)$ but with local non zero values $e_{p_{j}}$ equal to $\epsilon_{1}\hat{w_{j}}$. For each perturbation vector $\mathbf{e_{p}}$, the residual evaluation in Eqs. (\ref{fulldiscHessian}) and (\ref{fulldiscHessian2}) should be done for all the $\mathbf{e_{p}}^{\prime}$ and then summed as in Eq. (\ref{Hprime}) to obtain the column of $\mathbf{H}^{\prime}$ given by the considered $\mathbf{e_{p}}$.The total cost of the method would thus raise to $n_{res}=\left( n_{c}n_{s}^{d}\right)^{2}$ residual evaluations.

Finally, we also imposed $\epsilon_{2}$ to be of the form $\epsilon_{2}=\epsilon_{m_{2}}\left(\left|w\right|+1\right)$, forthcoming results will show that the choice of $\epsilon_{m_{2}}$ appeared to be more complex as the gradients are more sensitive to this choice. In particular, several values of local $\epsilon_{m_{2}}$
adapted to each flow variable were tested to obtain the best epsilon set (see \S \ref{sec5}).

\subsection{Intrusive method}

Optimal design methods require the evaluation of aerodynamic quantities with respect to some parametrization of the flow \cite{Sobieszczanski1987}. The solution is obtained using a gradient based optimization process which requires the computation of the product $\mathbf{J}\mathbf{a}$ where $\mathbf{a}$ is a specific vector field. Usually, $\mathbf{J}$ is obtained using an analytical linearization rather than a discrete linearization for precision purpose, as the optimization process is very sensitive to the precision of the Jacobian and its adjoint \cite{Peter2010}. However, due to the complexity of the equations to linearize, several simplifications may be done in the linearization process. For example, the thin layer assumption \cite{Candler1987} may be used so that cross derivatives of the stress tensors in the RANS equations are neglected.

Despite the simplifications achieved in the linearization, such optimization codes can be used to compute the sensitivity gradients. Indeed, the code can be intrusively modified in order to yield a product $\mathbf{J}\mathbf{u}$ for any vector $\mathbf{u}$. Using the same set of vector $\left(\mathbf{e_{p}}\right)$ as before, we can obtain all the Jacobian terms by matrix vector evaluations.

The matrix $\mathbf{H}^{\prime}$ can then be obtained using finite differences applied directly to the Jacobian. Indeed we have from Eq. (\ref{dJv2}):
\begin{equation} \label{opt}
\dfrac{\mathbf{J}_{\mathbf{w_{b}}+\epsilon_{2}\mathbf{e_{p}}} \hat{\mathbf{w}} - \mathbf{J}_{\mathbf{w_{b}}} \hat{\mathbf{w}}}{\epsilon_{2}}={\mathbf H}'\mathbf{e_{p}}.
\end{equation}
For each vector $\mathbf{e_{p}}$, the Jacobian associated to the perturbed baseflow $\mathbf{w_{b}}+\epsilon_{2}\mathbf{e_{p}}$ is obtained using the above mentioned Jacobian computation. Subtracting it with the unperturbed baseflow Jacobian and multiplying by the global mode $\hat{\mathbf{w}}$ we obtain ${\mathbf H}'\mathbf{e_{p}}$ so that ${\mathbf H}'$ can be formed explicitly using Eq. (\ref{opt}). We shall note that if no approximations are done in the analytical linearization, such a procedure would be more precise then our fully discrete approach since only a first order derivative would be approximated using finite differences.

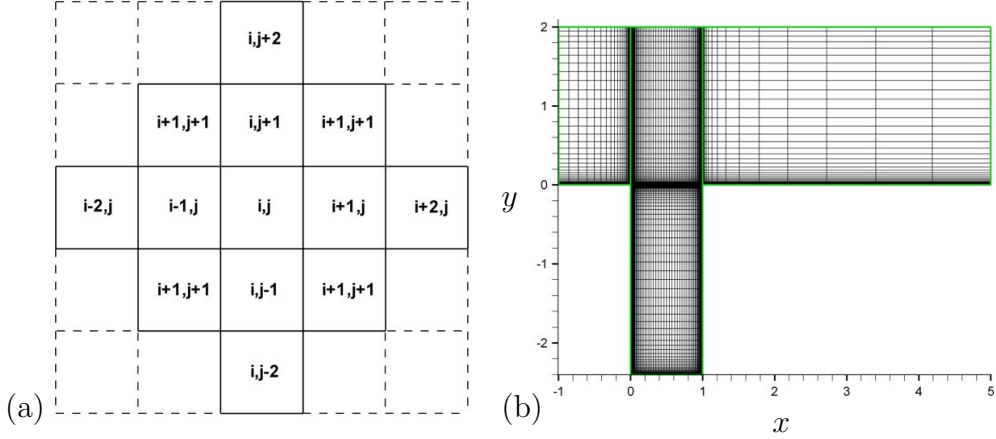
\begin{figure}
  \setlength{\unitlength}{1cm}
  \begin{picture}(8,3.25)
    \put(0,0){\includegraphics*[width=0.5\textwidth]{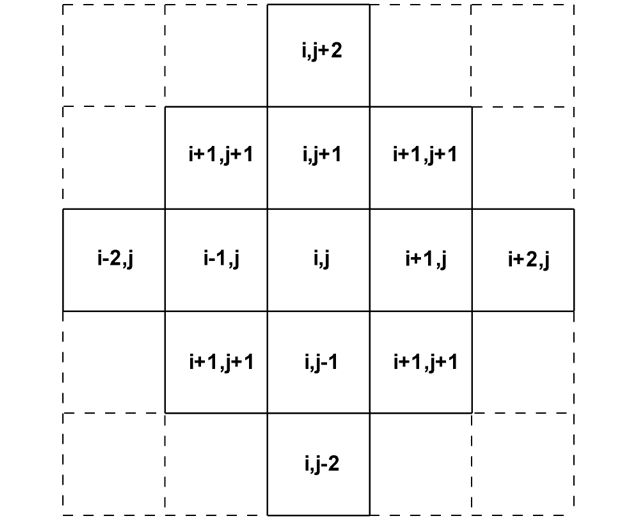}}
    \put(0,0){(a)}
    \put(6.6,0){\includegraphics*[width=0.5\textwidth]{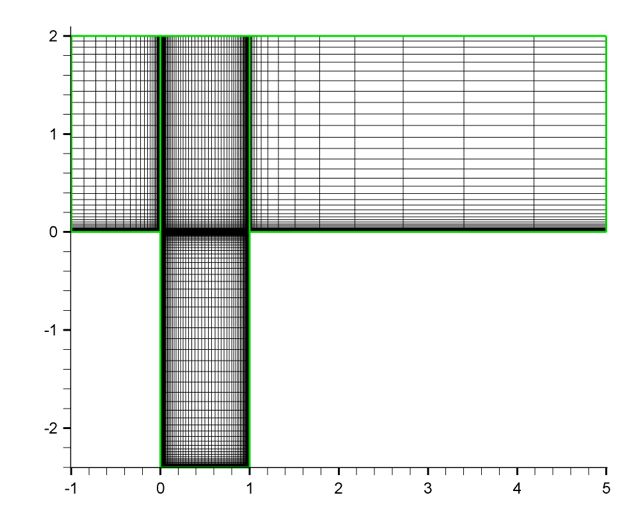}}
    \put(6.6,0){(b)}
    \put(10.2,-0.2){$x$}
    \put(6.6,2.8){$y$}
  \end{picture}
  \caption{(a): Example of stencil dependency of the residuals evaluated at the point $\left(i,j\right)$. (b): Mesh discretization example.} \label{fig1}
\end{figure}

\section{Numerical experiments} \label{sec4}

\subsection{Discretization of the flow equations}

The discretized Navier-Stokes equations in (\ref{eq1}) can be rewritten as:
\begin{equation} \label{eq1split}
\dfrac{d}{dt}\begin{pmatrix}
\mathbf{w}^{\text{mf}} \\
\mathbf{w}^{\text{tf}}
\end{pmatrix}
=
\begin{pmatrix}
\cal{R}^{\text{c,mf}}+\cal{R}^{\text{d,mf}} \\
\cal{R}^{\text{c,tf}}+\cal{R}^{\text{d,tf}} + \cal{T}
\end{pmatrix}
\end{equation}
where the superscripts $mf$ and $tf$ refer respectively to the mean and turbulent fields of the RANS equations. In particular,
$\mathbf{w}^{\text{mf}}=\left( \rho,\rho\mathbf{U},\rho E\right)^{T}$ where $\rho$ designates the density , $\mathbf{U}$ the velocity and $E$ the kinetic energy of the flow. Terms $\cal{R}^{\text{c}}$, $\cal{R}^{\text{d}}$ and $\cal{T}$ correspond respectively to the convective and diffusive fluxes of the equations and the turbulence source term.

Two turbulence models both widely encountered in practical CFD simulations are used. On the one hand, the $k-\omega$ model of Wilcox  \cite{Wilcox1988} which involves two turbulent variables with  $\mathbf{w}^{\text{tf}}=\left(\rho k ,\rho\omega\right)^{T} $, where $k$ is the turbulent kinetic energy and $\omega$ the rate of dissipation of turbulence. On the other hand, the turbulence model of Spalart and Allmaras \cite{Spalart1994}, a one equation turbulence model which involves the kinematic viscosity transform  $\tilde{\nu}$ with $\mathbf{w}^{\text{tf}}=\left(\rho\tilde{\nu}\right)$. The complete definition of the full set of equations for both turbulence models are detailed in their continuous form in \ref{ap1}.

We use the finite volume code elsA developed at ONERA \cite{Cambier2012} both to extract the baseflow and perform the residual evaluations (required for the Jacobian computation) on a two dimensional structured mesh. Steady state solutions are obtained using a backward-Euler scheme with local time-stepping.
In order to check the robustness of the method, several spatial discretization schemes of the mean field  convective fluxes $\cal{R}^{\textbf{c,mf}}$ were tested for the Jacobian computation: a central difference formula with Jameson's scalar dissipation and Martinelli's correction \cite{Martinelli1987}, a Roe scheme extended to the second order using MUSCL method \cite{VanLeer1979} and an AUSM scheme \cite{Mary2000}. The convective fluxes associated to the turbulence equations $\cal{R}^{\textbf{c,tf}}$ are discretised using the first order Roe scheme with Harten's correction to prevent the occurrence of low eigenvalues \cite{Harten1983}.
A central difference scheme is used for the turbulent diffusive fluxes. The viscous flux of the mean field is calculated at the interface by averaging cell-centered values of flux density which is computed from cell-centered evaluation of gradients. The source terms are discretized using estimates of gradients and variables at cell centers. The Zheng limiter operator \cite{Zheng1997} (which is designed to limit the values of $\rho\omega$) is used for the baseflow computation with the $k-\omega$ model, but is switched off for the stability analysis. These discretization choices all lead to an $n_{s}=5$ points stencil, an example of the dependency of the residual evaluated at the cell $\left(i,j\right)$ being depicted in Fig.\ref{fig1}(a). As mentioned in \S \ref{sec3}, the total number of points $n_{p}=13$ contributing to the residual evaluation at one point does not scale with $n_{s}^{2}=25$ (in a three dimensional case we would have $n_{p}=34$ rather than $n_{s}^{3}=125$). Boundary conditions are imposed by computing the residuals at the interfaces defined by the boundaries. The characteristic equations are integrated to obtain boundary values in the case of inlet or outlet conditions. Turbulent quantities at walls are computed as proposed by Wilcox \cite{Wilcox1988} and Spalart and Allmaras \cite{Spalart1992}.
 Note that all these discretization choices combined with turbulence model yield to second order differentiable discrete equations as required for the sensitivity gradients to be defined.

The elsA software includes a shape optimization module in which the discrete RANS equations and various turbulence models were analytically linearized \cite{Peter2006,Din2006,Peter2012,Renac2011}. We modified this code as stated in \S \ref{sec3} in order to enable the computation of both matrices $\mathbf{J}$ and $\mathbf{H}^{\prime}$, for the turbulence model $k-\omega$ model of Wilcox and the Roe scheme for the mean field convective fluxes, other terms being discretized as described above. Nonetheless, the analytical linearization in the module was done using the thin layer assumption \cite{Candler1987}, so that we expect to observe some differences when comparing results obtained with this strategy to those obtained by the fully discrete approach.

\subsection{Description of the test-case}

As an application case, we consider a two dimensional cavity of height $D=0.12$m and width $L=0.05$m ($L/D=0.42$) as illustrated in Fig.\ref{fig1}(b). The flow is compressible with a Mach number of 0.8, stagnation conditions being equal to $94 400$Pa for the pressure and $292.5$K for the temperature. The Reynolds number based on the free stream velocity $U_{\infty}$, density $\rho_{\infty}$, temperature $T_{\infty}$, the cavity length $L$ and Sutherland's law for the viscosity is equal to 860 000. We impose a turbulent parallel profile with a boundary layer thickness $\delta\text{=2.3mm}$ at the inlet of the domain.
The lower part of the domain is composed of an adiabatic wall, while a wall slip condition is imposed on the upper part of the domain, the outlet static pressure $p_{\infty}$ being fixed at $61 900$Pa. All quantities are nondimensionalized using the free stream variables $\rho_{\infty}$,$U_{\infty}$,$T_{\infty}$ and the cavity length $L$.

The mesh used for the simulations is depicted in Fig.\ref{fig1}(b) and is composed of three vertical blocs.  For each block, we either use a tangential or semi-tangential law for the evolution of the discretization step. The cells adjacent to the cavity corners are squares of size $\Delta x=7.0E-05$ imposing $\Delta y^{+} =1.4$ on the upstream wall, ensuring that the first discretization points are inside the viscous sublayer. The discretizations associated to each bloc are summarized in Table \ref{tablemesh}, yielding a total number of cells $N_{m}=295 000$.

\begin{table}[b]
\centering       
\setlength{\tabcolsep}{1pt}
\begin{tabular}{c|c|c|c|c}
 {}   &        $x$       &         $y $       & Discretization points & Number of cells\\
\hline
Bloc 1 & $-1\le x_{1}\le 0$ &  \quad$0\le y_{1}\le 2$ & $151\times 221$ & 33000 \\
Bloc 2 &  \quad$0\le x_{2}\le 1$  & $-2.4\le y_{2}\le 2$& $401\times 601$ & 240000\\
Bloc 3 &  \quad$1\le x_{3}\le 5$ &  \quad$0\le y_{3}\le 2 $& $101\times 221 $& 22000
\end{tabular}
\caption{\label{tablemesh} Definition of the blocs and their discretization properties.}
\end{table}

\subsection{Baseflow computation}

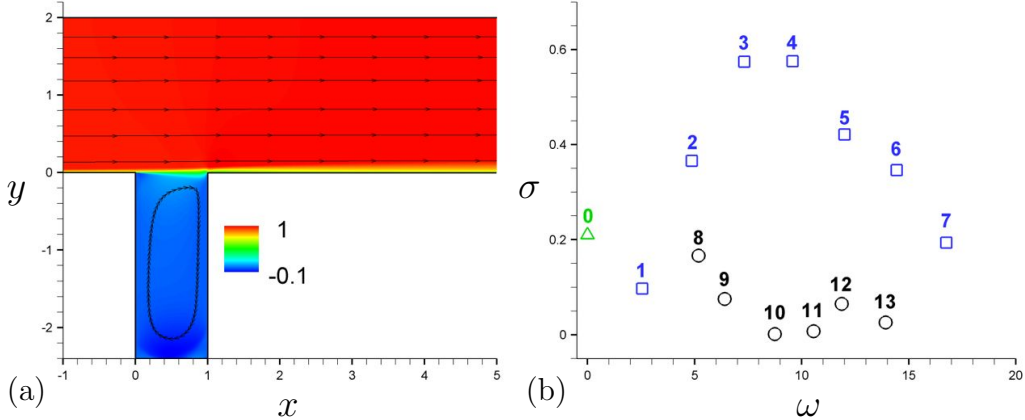
\begin{figure}[h!]
\setlength{\unitlength}{1cm}
\begin{picture}(8,6)
    \put(0.1,.2){\includegraphics*[width=0.5\textwidth]{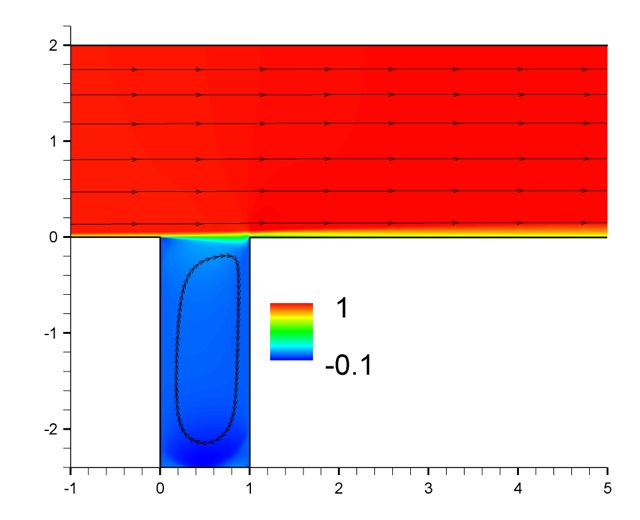}}
    \put(0.1,.2){(a)}
    \put(3.7,0){\large{$x$}}
    \put(0.1,2.9){\large{$y$}}
    \put(7,0.2){\includegraphics*[width=0.5\textwidth]{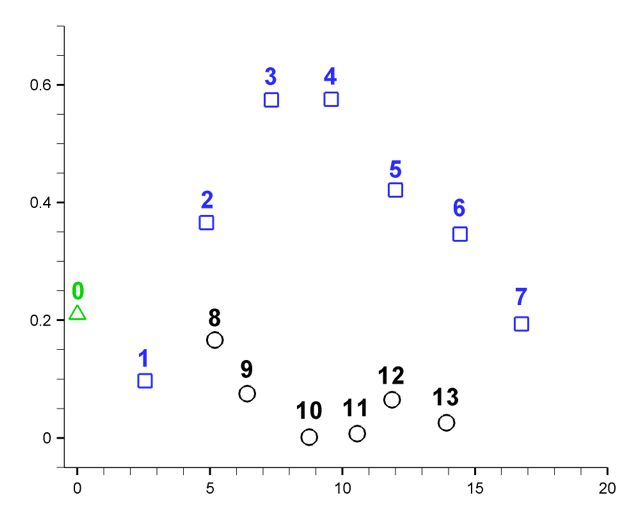}}
    \put(7,0.2){(b)}
    \put(10.6,0){\large{$\omega$}}
    \put(6.9,2.9){\large{$\sigma$}}
  \end{picture}
\caption{(a): Baseflow streamlines and streamwise velocity contours. (b): Unstable eigenvalues obtained using the $k-\omega$ model of Wilcox (Roe scheme).}   \label{figspec}
\end{figure}

Convergence of the baseflow is assessed by ensuring that the explicit residuals of the mean field equations are small (typically $10^{\text{-11}}$) and that the residual of the turbulent equations have decreased by several orders of magnitude. Streamlines and streamwise velocity component of the baseflow $\mathbf{w_{b}}$ obtained with the Roe scheme and the $k-\omega$ model of Wilcox are plotted in Fig.\ref{figspec}(a). We observe the formation of a mixing layer induced by the presence of a large recirculation bubble inside the cavity and growing from the upstream corner of the cavity.
This configuration corresponds to the experimental study of \citet{Forestier2003} who characterized this flow to be unsteady with a dominant frequency around 2000 Hz. The mixing layer is subject to Kelvin-Helmholtz instabilities which lead to the creation of vortices that impact the downstream corner of the cavity. This impact generates pressure waves propagating upstream that perturb the mixing layer sustaining the instability mechanism. This mechanism of aeroacoustic feedback was proposed by Rossiter \cite{Rossiter1966}. Note that the unsteady RANS simulations recover accurately the frequency selection of the natural flow \cite{Lawson2011}.

In order to compare the results obtained with the different turbulence models and discretization schemes regardless of their dependence to the baseflow, we chose to keep the depicted baseflow  for all stability computations (regardless of the equations to linearize for the stability analysis). A conversion function \cite{Deck2011} is applied to compute $\rho\tilde{\nu}$ for the Spalart-Allmaras model from $\rho k$ and $\rho\omega$ by matching the eddy viscosity of both models.

\begin{table}[h]
\centering       
\setlength{\tabcolsep}{1pt}
\begin{tabular}{c|c|c|c|c|c|c|c|c}
 Model           & $n_{c}$ &  $N$ & $M_{v}$ & $n_{res}$ &    $n_{e}$  &$n_{e}^{obt}$ &  $n_{e}^{obt}/N^{2}$ & $M_{J}$ \\
\hline
Spalart-Allmaras &  5      & $1475.10^{3}$ & 11 MB& 125 & $96.10^{6}$ & $63.10^{6}$ & $3.10^{-05}$ & 0.9 GB\\
$k-\omega $      &  6      & $1770.10^{3}$ & 14 MB &150 & $138.10^{6}$& $80.10^{6}$ & $3.10^{-05}$ & 1.2 GB
\end{tabular}
\caption{\label{tablejac} Jacobian matrix dimensions. }
\end{table}

\subsection{Memory cost of Jacobian computation, storage and inversion}

The Jacobian matrix is extracted with the method presented in \S \ref{sec3} and stored on disk. The method requires $n_{res}=25n_{c}$ residual evaluations and each vector of size $N$ shall be stored. The quantities characterising the Jacobian size for both $k-\omega$ and Spalart-Allmaras turbulence models are summarized in Table \ref{tablejac}, where we introduce $M_{v}$ and $M_{J}$ the memory costs of storing a real vector and the Jacobian matrix respectively.

We can remark that the obtained number of non zero elements $n_{e}^{obt}$ is about 30 percent lower than the maximum potential non zero elements $n_{e}$ introduced in \S  \ref{sec3}. This is not surprising as all conservative variables do not intervene in each equation. The matrices are sparse with very small ratio of non zero elements to their size $n_{e}^{obt}/N^{2}$.

\begin{table}[h]
\centering       
\setlength{\tabcolsep}{1pt}
\begin{tabular}{c|c|c|c|c}
 Model           & Max memory &  Number of procs & Memory per proc  & Time per proc \\
\hline
Spalart-Allmaras &  52 GB      & 24 & 2.1 GB & 214 s\\
$k-\omega $      &  65 GB      & 24 & 2.7 GB & 256 s
\end{tabular}
\caption{\label{tablecost} Jacobian matrix inversion cost.}
\end{table}

The eigenvalue problem in Eq. (\ref{linear1}) is then solved using a shift and invert strategy with direct inversion of the matrix as described in \S \ref{sec2}. Direct inversions are fast and accurate but require large amount of memory. We show in Table \ref{tablecost} the total computational cost of one direct inversion of a complex matrix (we use complex shifts to focus on some particular eigenvalues) in terms of maximum amount of memory, number of processors and computational time per processors.

The maximum memory is reached during the LU factorisation of the matrix and is about 50 times the matrix size, the inversions being quickly processed. We foresee that the increase of memory would become prohibitive for very large systems ($n_{e}^{obt}> 10^{9}$). Note that the scope of this study is not to propose an optimal method in terms of computational time or memory cost to compute the sensitivity gradients, but lies in the discrete definition and computation of these quantities. However, the method overview presented in \S \ref{sec2} presents a fully on-the-fly approach for optimization of this procedure.

\section{Results} \label{sec5}

\subsection{Linear stability analysis}

\subsubsection{Unstable modes}

We first consider the results obtained with the $k-\omega$ model of Wilcox and the Roe scheme for the convective flux discretization. Solving the eigenvalue problem in Eq. (\ref{linear1}) we obtain the set of unstable eigenvalues depicted in Fig.\ref{figspec}(b). We obtain a spectrum similar to the one computed by \citet{Yamouni2013} for a laminar compressible flow in a square cavity. We observe an upper branch (modes $1-7$, denoted with square symbols \textcolor{blue}{$\square$}) seemingly corresponding to Kelvin-Helmholtz modes and a lower branch (modes $8-13$, denoted with circle symbols \textcolor{black}{$\Circle$}) that we attribute to acoustic modes.

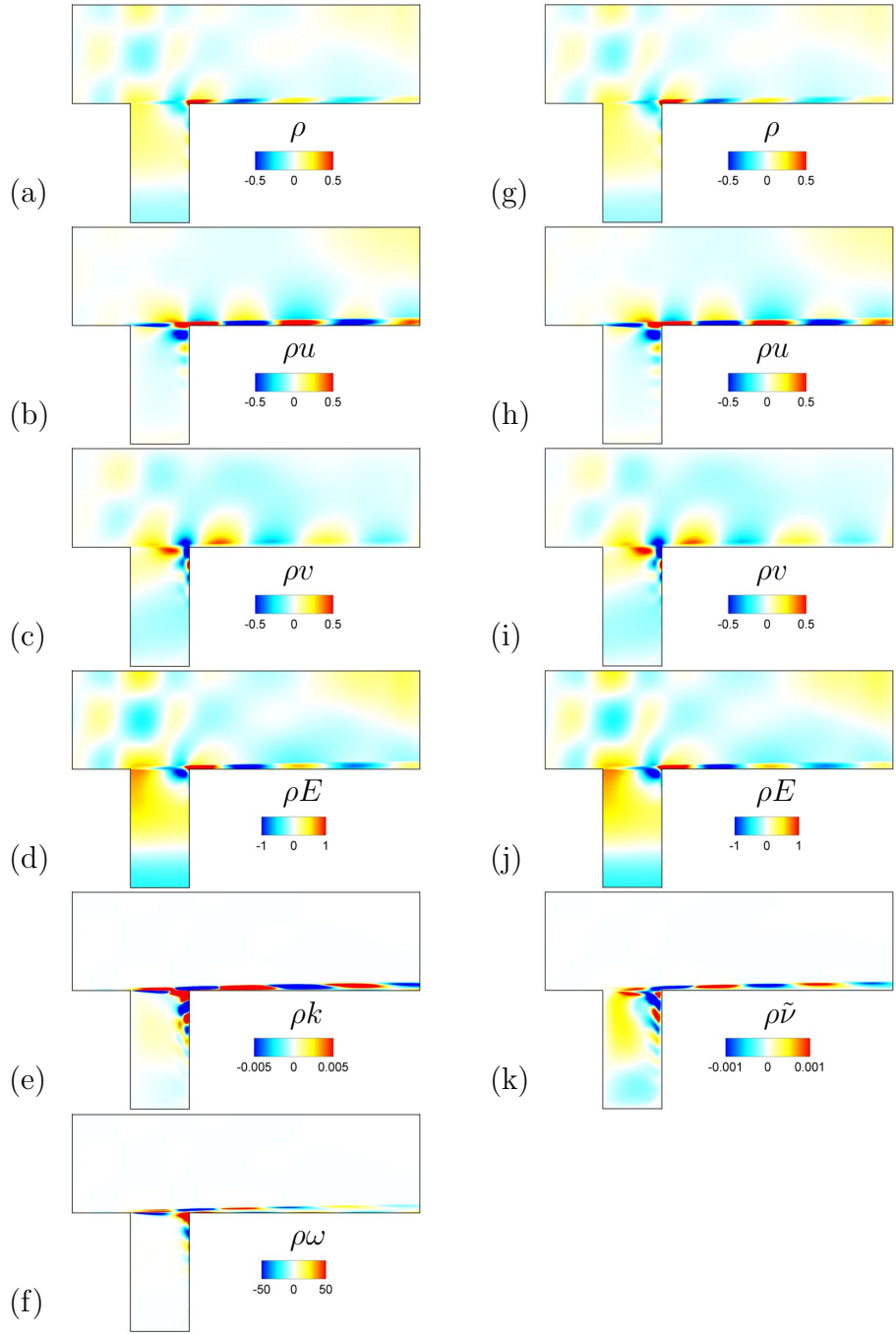
\begin{figure}
  \setlength{\unitlength}{1cm}
  \begin{picture}(10,18)
    \put(0.8,15){\includegraphics*[width=0.35\textwidth]{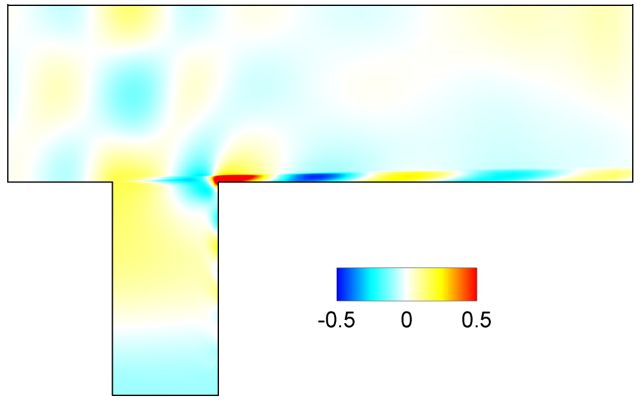}}
    \put(0,15.3){(a)}
    \put(0.8,12){\includegraphics*[width=0.35\textwidth]{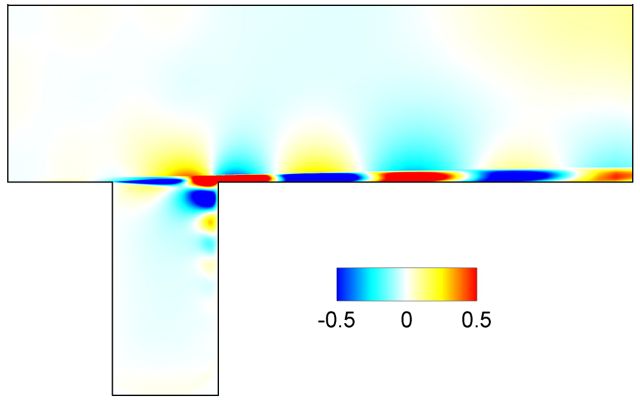}}
    \put(0,12.3){(b)}
    \put(0.8,9){\includegraphics*[width=0.35\textwidth]{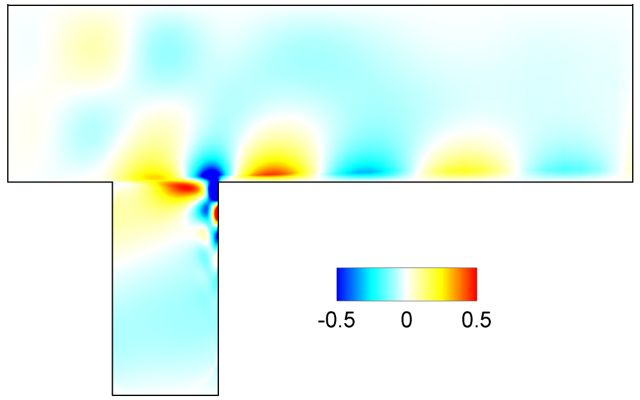}}
    \put(0,9.3){(c)}
    \put(0.8,6){\includegraphics*[width=0.35\textwidth]{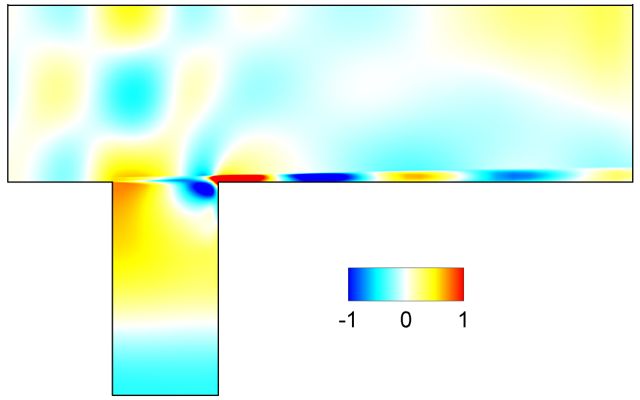}}
    \put(0,6.3){(d)}
    \put(0.8,3){\includegraphics*[width=0.35\textwidth]{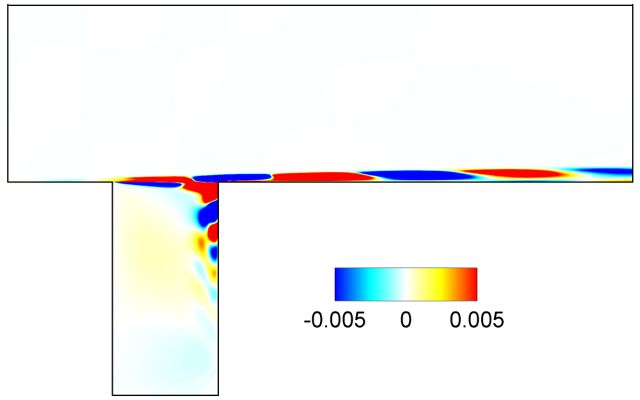}}
    \put(0,3.3){(e)}
    \put(0.8,0){\includegraphics*[width=0.35\textwidth]{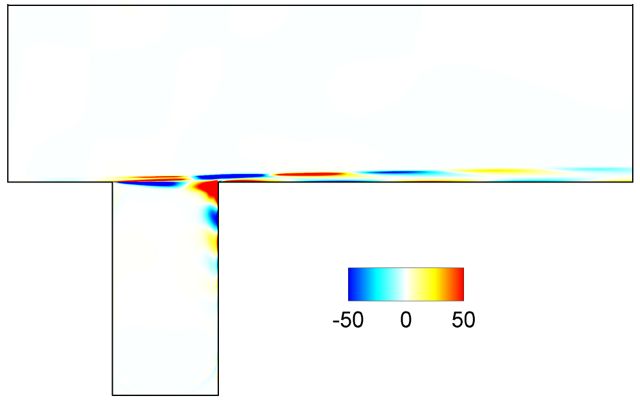}}
    \put(0,0.3){(f)}

    \put(3.8,16.2){$\rho$} 
    \put(3.7,13.2){$\rho u$}
    \put(3.7,10.2){$\rho v$}
    \put(3.7,7.2){$\rho E$}
    \put(3.8,4.2){$\rho k$}
    \put(3.8,1.2){$\rho\omega$}
    \put(7.2,15){\includegraphics*[width=0.35\textwidth]{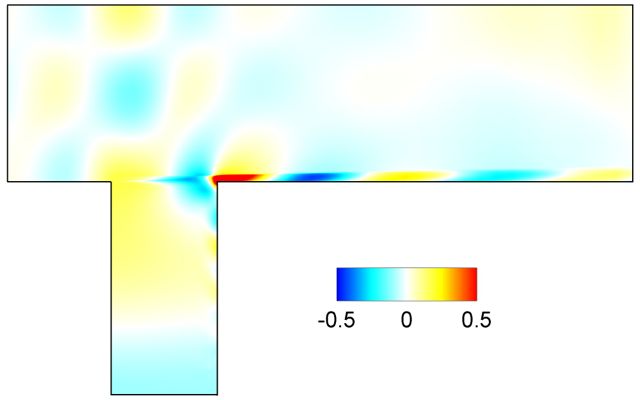}}
    \put(6.5,15.3){(g)}
    \put(7.2,12){\includegraphics*[width=0.35\textwidth]{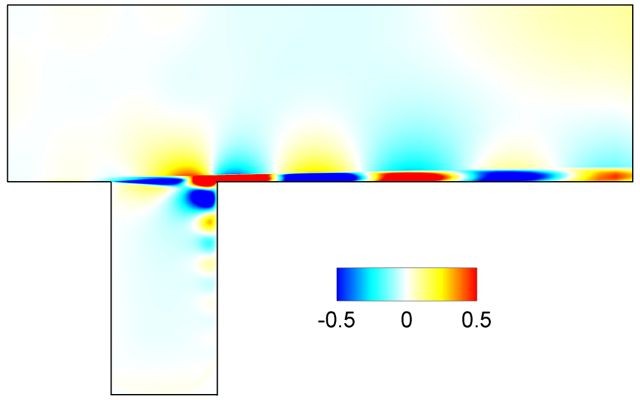}}
    \put(6.5,12.3){(h)}
    \put(7.2,9){\includegraphics*[width=0.35\textwidth]{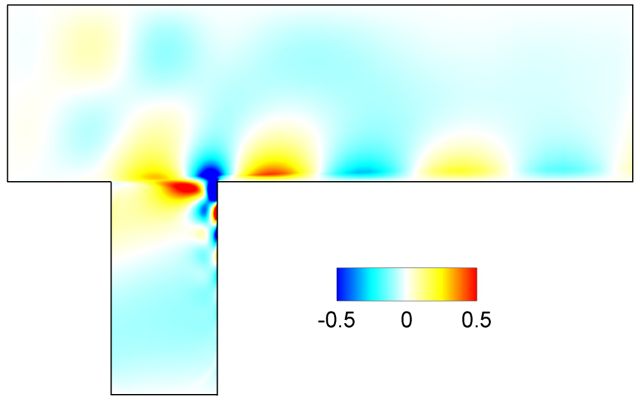}}
    \put(6.5,9.3){(i)}
    \put(7.2,6){\includegraphics*[width=0.35\textwidth]{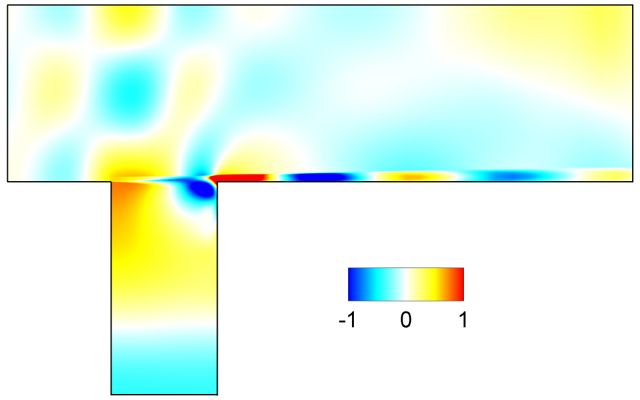}}
    \put(6.5,6.3){(j)}
    \put(7.2,3.0){\includegraphics*[width=0.35\textwidth]{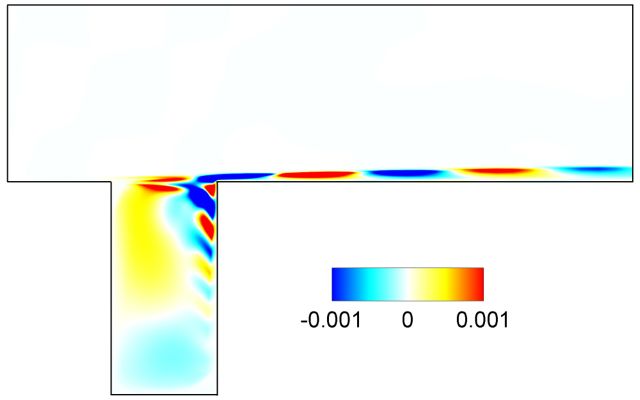}}
    \put(6.5,3.3){(k)}

    \put(10.2,16.2){$\rho$}
    \put(10.1,13.2){$\rho u$}
    \put(10.1,10.2){$\rho v$}
    \put(10.1,7.2){$\rho E$}
    \put(10.2,4.2){$\rho\tilde{\nu}$}
  \end{picture}
  \caption{Comparison of the spatial structure of the fundamental mode (mode $1$) obtained with the $k-\omega$ model of Wilcox (a,b,c,d,e,f) and the Spalart-Allmaras model (g,h,i,j,k). The real part of the different components are plotted.}
 \label{figmode}
\end{figure}

The upper branch is composed of the fundamental mode (mode $1$), which exhibits a frequency close to the natural flow frequency around 2000Hz ($\omega=2.4$), as well as several of its harmonics (modes $2-7$). These modes correspond to dynamical modes linked to the aeroacoustic feedback mechanism proposed by Rossiter \cite{Rossiter1966}. The fundamental mode is found to be dominant at large time when integrating the unsteady RANS equations (not shown here). We depicted in Figs.\ref{figmode}(a-f) the real part of its spatial structure.
Kelvin-Helmholtz instabilities grow from the upstream edge and propagate downstream. The turbulent fluctuations are located within the unstable Kelvin-Helmholtz like structures, with the downstream propagation of region of low and high values of turbulent kinetic energy ($\rho k$) and dissipation rate ($\rho\omega$). Note that in order to give sense to the comparison of modes scales, all modes are phased at $\left( x=-1,y=0\right)$ and normalized by setting the norm of their momentum equal to $1$.
The upper branch modes structures also present acoustic resonance patterns. As we consider compressible equations, acoustic resonance may occur in the cavity as suggested by East \cite{East1966}. The coupling between the aeroacoustic feedback and acoustic resonance mechanism was also studied by \citet{Yamouni2013}. They showed that the most unstable mode corresponds to an aeroacoustic mode for which acoustic resonance occurs.

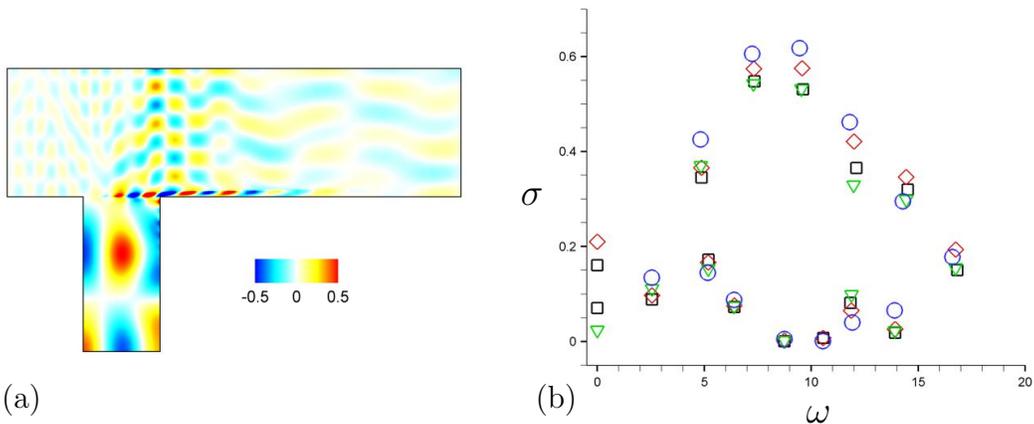
\begin{figure} [h]
  \setlength{\unitlength}{1cm}
  \begin{picture}(9,6)
    \put(0.1,.9){\includegraphics*[width=0.45\textwidth]{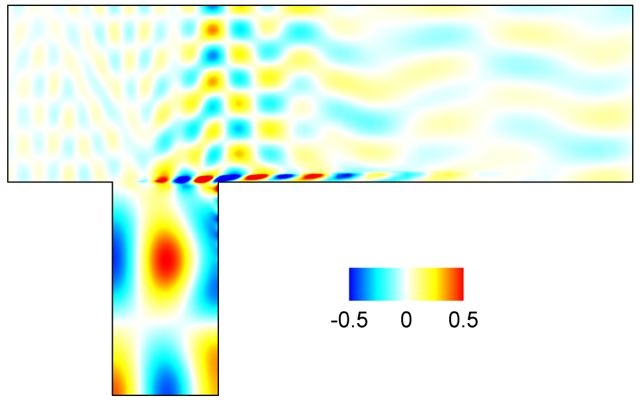}}
    \put(0.1,.2){(a)}
    \put(7.2,0.2){\includegraphics*[width=0.5\textwidth]{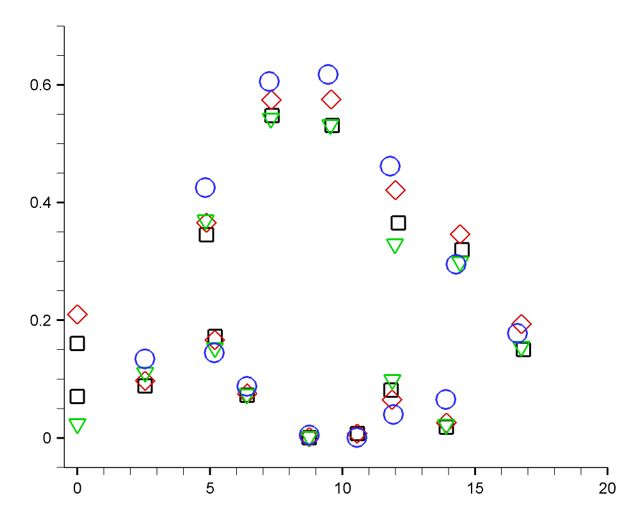}}
    \put(7.2,0.2){(b)}
    \put(10.8,0){\large{$\omega$}}
    \put(7.0,2.9){\large{$\sigma$}}
  \end{picture}
  \caption{(a): Real part of the $\rho$ component spatial structure for mode $10$. (b): Impact of the physical modelling on the spectrum: \textcolor{black}{$\square$} elsA optimization code, \textcolor{red}{\large{$\diamond$}} $k-\omega$ model of Wilcox, \textcolor{green}{$\triangledown$} Spalart-Allmaras model, \textcolor{blue}{$\Circle$} Uncoupled equations.}   \label{impact0}
\end{figure}

The lower branch (modes $8-13$) of unstable eigenvalues in Fig.\ref{figspec}(b) refers to unstable modes with smaller amplification rates and which exhibit strong patterns of acoustic resonance. As an example we plotted in Fig.\ref{impact0}(a) the spatial structure of the density for mode $10$. We clearly see stronger resonance patterns compared to mode $1$ in Fig.\ref{figmode}(a) (same scaling is used). These modes are likely to be acoustic resonance modes which became unstable under the excitation of Kelvin-Helmholtz instabilities. \\

\textbf{Remark:} The spatial structure of the unstable non-oscillating mode (mode $0$, denoted with a triangle symbol \textcolor{green}{$\triangledown$} in Fig.\ref{figspec}(b)) differs from the other modes. It is not located near the mixing layer but near the upstream wall of the cavity around $\left( 0,-0.4\right)$. As will be shown below, this mode is extremely sensitive to the numerical discretization and the turbulence modeling. These observations lead us to believe that it is a spurious mode.

\subsubsection{Validation of the numerical method}

The impact of the physical modelling is investigated using the Roe scheme with the baseflow obtained in \S \ref{sec4}. We plot in Fig.\ref{impact0}(b) the spectrum computed with the $k-\omega$ and Spalart-Allmaras turbulence models, with uncoupled equations and with the modified elsA optimization code. Uncoupled equations correspond to the mean field equations in Eq. (\ref{eq1split}) for which the turbulent viscosity is frozen in the linearization process, so that turbulent fluctuations are not considered \cite{Juan2006,Cossu2009,Hwang2010}.

A first interesting result is that the model choice (\textcolor{red}{\large{$\diamond$}}, \textcolor{green}{$\triangledown$} and \textcolor{blue}{$\Circle$} in  Fig.\ref{impact0}(b)) mainly affects the growth rate of the modes but not their frequency. This result is in agreement with Rossiter's mechanism where the frequency selection is only linked to the cavity width and Mach number. As for the amplification rate, we do observe some discrepancy between the intrusively modified elsA code (\textcolor{black}{$\square$}) and our fully discrete method (\textcolor{red}{\large{$\diamond$}}) suggesting that the thin layer assumption may have some impact on the spectrum in this configuration.

\begin{figure} [h!]
  \setlength{\unitlength}{1cm}
  \begin{picture}(8,4.5)
    \put(0.2,.4){\includegraphics*[width=0.45\textwidth]{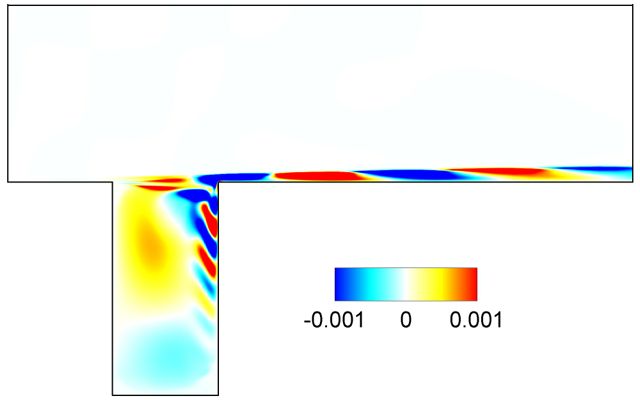}}
    \put(0.1,.2){(a)}
    \put(7,0.4){\includegraphics*[width=0.45\textwidth]{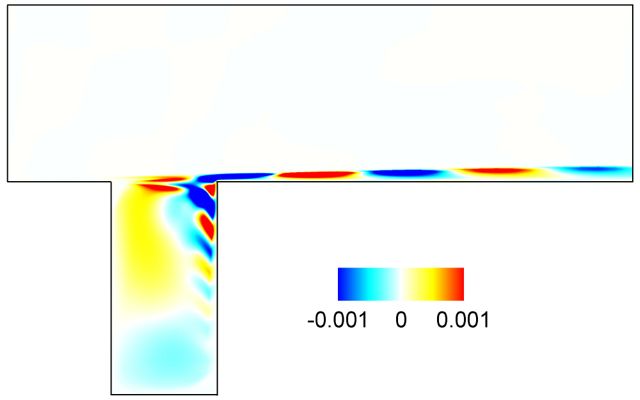}}
    \put(7,0.2){(b)}
  \end{picture}
  \caption{Eddy viscosity fluctutation $\mu_{t}^{\prime}$ induced by mode $1$. (a): $k-\omega$ model of Wilcox. (b): Splalart-Allmaras model.}   \label{f:mut}
\end{figure}

The modelling does not have a strong impact on the modes although some tendency can be observed. Uncoupling the equations seems to increase most unstable modes growth rate suggesting that the discarded term representing eddy viscosity fluctuations $\mu_{t}^{\prime}$ is likely to dissipate some energy. On the contrary, the Spalart-Allmaras modes seem to be more dissipative with smaller growth rates. The cavity modes ($8-13$) are less affected by the physical modelling as they correspond to acoustic resonance mode that are more inviscid in nature.

The spatial structure of the Spalart-Allmaras fundamental mode is compared to the $k-\omega$ mode in Figs.\ref{figmode}(g-k). We observe strong similarities between both modes structures for the mean field variables. In order to compare the relative contributions of the different components fluctuations to the baseflow, we summarize in Table \ref{tableComp} for each conservative variable the ratio of the mode maximum value to the baseflow maximum value (here for mode $1$). This ratio being defined up to an arbitrary amplitude, we rescale it by setting the variable $\rho$ ratio to $1$ for both modes.
\begin{table}[h]
\centering       
\setlength{\tabcolsep}{1pt}
\begin{tabular}{c|c|c|c|c|c|c|c|c}
 Model           & $\rho$ &  $\rho u$ & $\rho v$ & $\rho E$ & $\rho k$ & $\rho\omega$ &  $\rho\tilde{\nu}$ & $\mu_{t}$  \\
\hline
Spalart-Allmaras &  1      & 5 & 2& 1 & . & .  & 0.8 & 0.8 \\
$k-\omega $      &  1      & 6 & 2& 1 & 13 & 48& . & 1.5
\end{tabular}
\caption{\label{tableComp} Ratio for each conservative variable of the maximum value of mode $1$ to the maximum value of the baseflow.}
\end{table}
We observe that the turbulent fluctuations obtained using both turbulence models strongly impact the baseflow compared to the mean field variables, suggesting that the turbulent quantities do seem to play a role in the instability mechanism. In order to compare the impact of both turbulence models, we compute the eddy viscosity fluctuation $\mu_{t}^{\prime}$ associated with the mode fluctuations, derivations of $\mu_{t}^{\prime}$ for both models being detailed in \ref{ap2}. We can observe in Fig.\ref{f:mut} that both turbulence models lead to very similar fluctuation fields in terms of structure and order of magnitude.

\begin{figure} [h!]
  \setlength{\unitlength}{1cm}
  \begin{picture}(8,6)
    \put(0.2,.2){\includegraphics*[width=0.5\textwidth]{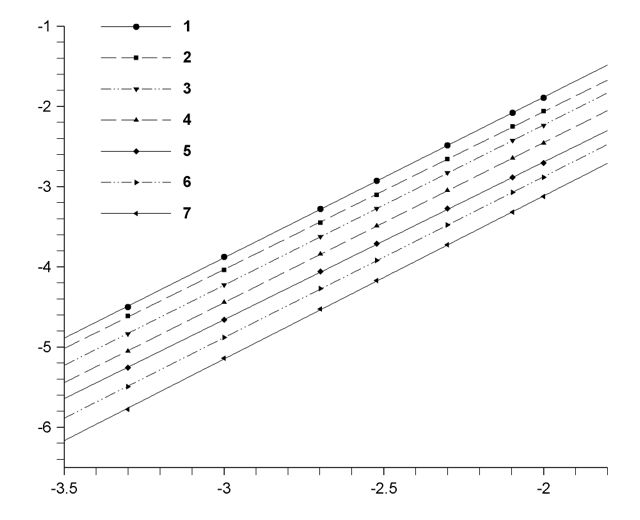}}
    \put(0.1,.2){(a)}
    \put(3.25,0){$\log_{10}\left(\epsilon_{m}\right)$}
    \put(0.1,5.8){$\log_{10}\left(err\right)$}
    \put(7,0.2){\includegraphics*[width=0.5\textwidth]{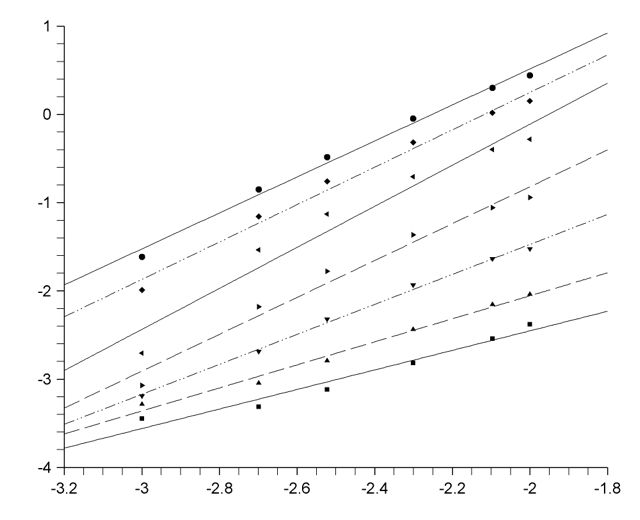}}
    \put(7,0.2){(b)}
    \put(10.15,0){$\log_{10}\left(\epsilon_{m}\right)$}
  \end{picture}
  \caption{Convergence with $\epsilon_{m}$ of the upper branch unstable eigenvalue ($1-7$) for (a): the $k-\omega$ model of Wilcox, (b): the Splalart-Allmaras model.}   \label{order2}
\end{figure}
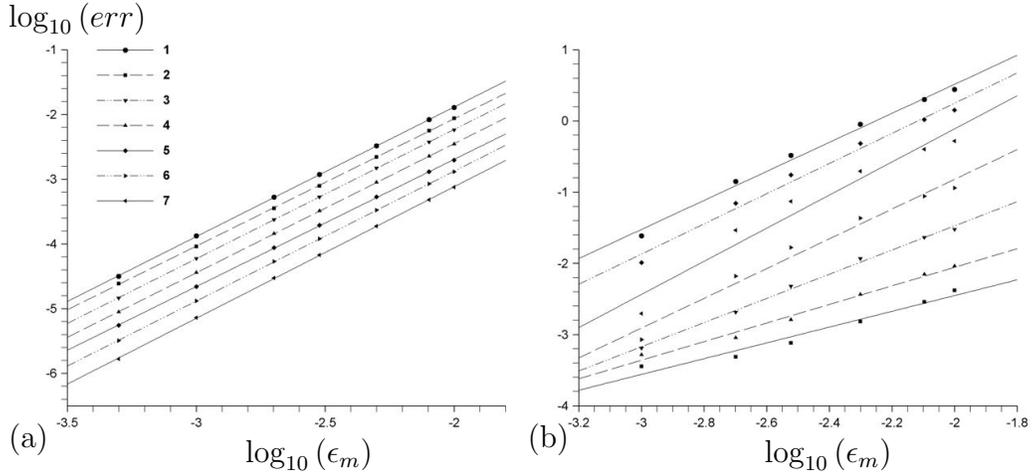

In order to check the convergence of the method as a function of $ \epsilon_m $, we extracted the set of eigenvalues $\lambda_{\epsilon_{m}}$ for Jacobian matrices computed with various values of $\epsilon_{m}$. The spectrum is found converged for $\epsilon_{m}<10^{-5}$, we thus use as a reference the set of eigenvalues $\lambda_{0}$ computed for $\epsilon_{m}=5.10^{-6}$. We then compute the relative error $err=\left|\lambda_{\epsilon_{m}}-\lambda_{0}\right|/\left|\lambda_{0}\right|$ with $\epsilon_{m}$. We plot in Fig.\ref{order2} the base $10$ logarithm of these quantities for the dynamical branch eigenvalues (modes $1-7$) and interpolate the different sets with linear fits. Note that the curves were arbitrarily shifted from each other to ease visualisation. The slopes $a$ obtained with the linear fit evaluation as well as the regression parameter $R^{2}$ are summarized in Table \ref{tableconv} for the different modes and both turbulence models. We observe a strong convergence of the method for the $k-\omega$ model of Wilcox with a slope coefficient of $2$ for nearly all the modes, the convergence coefficients for the Spalart-Allmaras modes being lower but still greater than $1$.

\begin{figure} [h]
  \setlength{\unitlength}{1cm}
  \begin{picture}(8,6)
    \put(0.2,.2){\includegraphics*[width=0.5\textwidth]{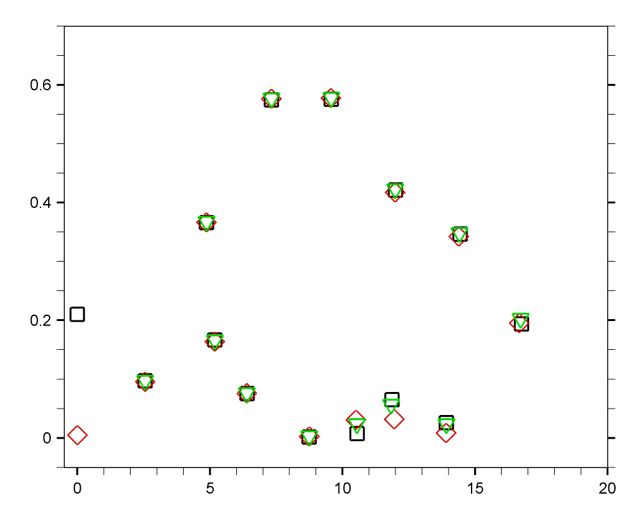}}
    \put(0.1,.2){(a)}
    \put(3.7,0){\large{$\omega$}}
    \put(0.1,2.9){\large{$\sigma$}}
    \put(7,0.2){\includegraphics*[width=0.5\textwidth]{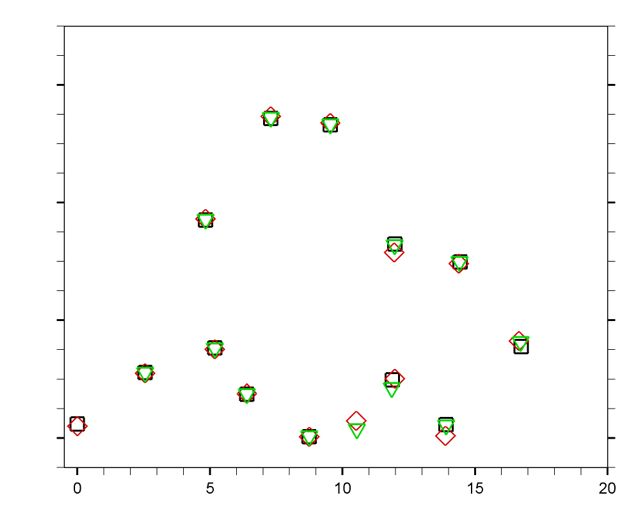}}
    \put(7,0.2){(b)}
    \put(10.6,0){\large{$\omega$}}
  \end{picture}
  \caption{Influence of the discretization scheme on the spectrum, \textcolor{black}{$\square$} Roe scheme, \textcolor{red}{\large{$\diamond$}} Jameson scheme, \textcolor{green}{$\triangledown$} AUMS scheme. (a): $k-\omega$ model of Wilcox. (b): Spalart-Allmaras model.} \label{impact}
\end{figure}
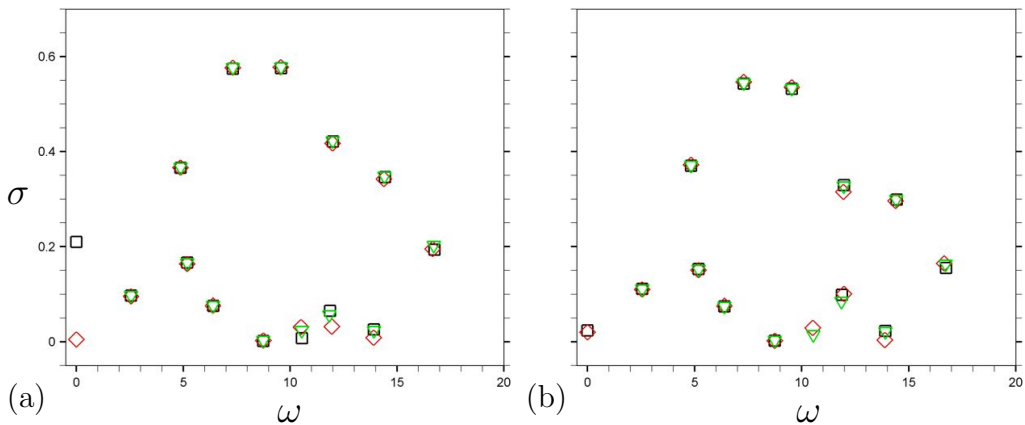

\begin{table}[h]
\centering       
\setlength{\tabcolsep}{1pt}
\begin{tabular}{c|c|c|c|c|c|c|c|c|c|c|c|c|c|c}
 Mode number           & \multicolumn{2}{|c}{$1$} & \multicolumn{2}{|c}{$2$} & \multicolumn{2}{|c}{$3$} & \multicolumn{2}{|c}{$4$} & \multicolumn{2}{|c}{$5$} & \multicolumn{2}{|c}{$6$} & \multicolumn{2}{|c}{$7$} \\
\hline
Parameters &  $a$ & $R^{2}$  & $a$ & $R^{2}$ & $a$ & $R^{2}$ & $a$ & $R^{2}$ &  $a$ & $R^{2}$ & $a$ & $R^{2}$ & $a$ & $R^{2}$  \\
\hline
Spalart-Allmaras & \hspace{0.1cm} $1.11$\hspace{0.1cm}&0.96 & 1.31&0.99 & 1.70&0.99 & 2.10&0.97 &  2.33&0.95 & 2.12&0.98 & 2.04&0.99  \\
\hline
$k-\omega $      & \hspace{0.1cm} $2.00$\hspace{0.1cm}&0.99 & 1.97&0.99 & 1.99&0.99 & 1.99&0.99 &  1.97&0.99 & 2.00&0.99 & 2.02&0.99  \\
\end{tabular}
\caption{\label{tableconv} Linear fit parameters of the eigenvalues convergence with $\epsilon_{m}$.}
\end{table}

The impact of the numerical scheme was investigated for both turbulence models using the different schemes presented in \S \ref{sec4}. Results obtained are depicted in Fig.\ref{impact} for (a) the $k-\omega$ model of Wilcox and (b) the Spalart-Allmaras model. As expected, the spectrum is poorly affected by the choice of numerical discretization, especially for the Kelvin-Helmholtz branch ($1-7$).

\begin{figure}[h!]
  \setlength{\unitlength}{1cm}
  \begin{picture}(10,9)
    \put(0.8,6){\includegraphics*[width=0.35\textwidth]{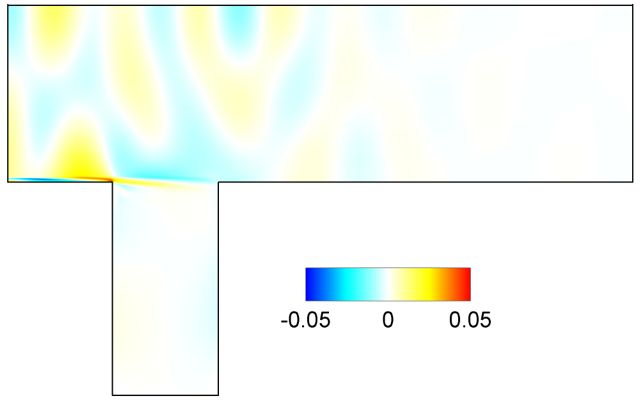}}
    \put(0,6.3){(a)}
    \put(0.8,3){\includegraphics*[width=0.35\textwidth]{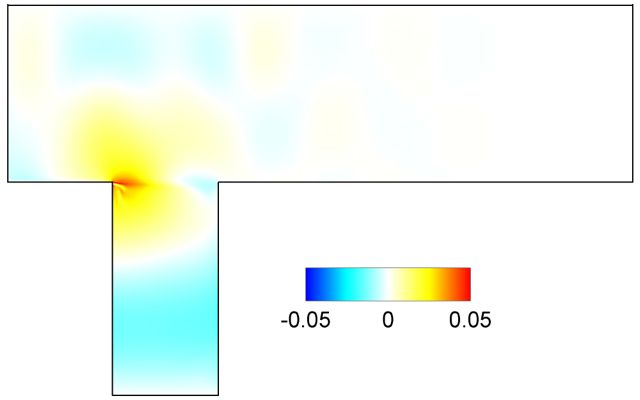}}
    \put(0,3.3){(c)}
    \put(0.8,0){\includegraphics*[width=0.35\textwidth]{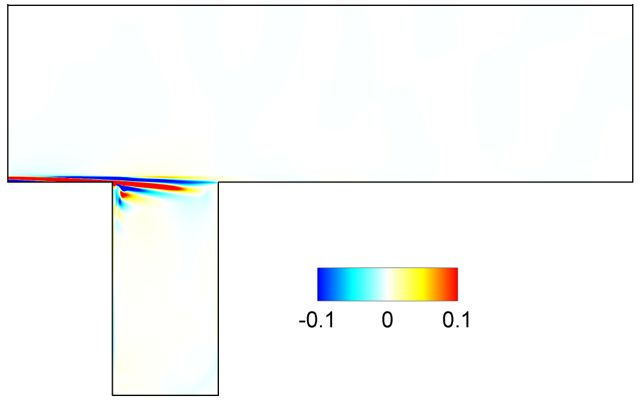}}
    \put(0,0.3){(e)}
    \put(3.6,7.2){$\rho$}
    \put(3.5,4.2){$\rho v$}
    \put(3.5,1.2){$\rho k$}
    \put(7.2,6){\includegraphics*[width=0.35\textwidth]{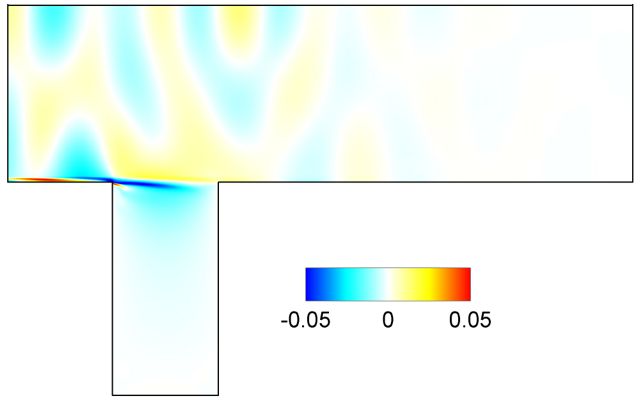}}
    \put(6.5,6.3){(b)}
    \put(7.2,3){\includegraphics*[width=0.35\textwidth]{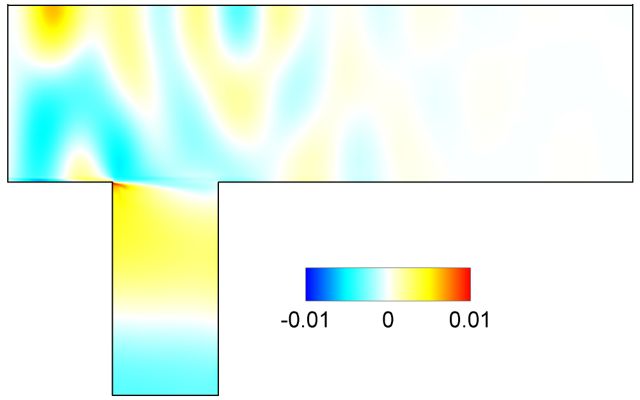}}
    \put(6.5,3.3){(d)}
    \put(7.2,0.0){\includegraphics*[width=0.35\textwidth]{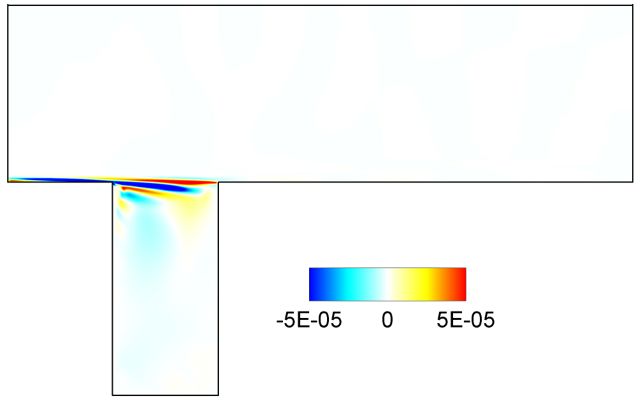}}
    \put(6.5,0.3){(f)}
    \put(9.9,7.2){$\rho u$}
    \put(9.9,4.2){$\rho E$}
    \put(9.9,1.2){$\rho\omega$}
  \end{picture}
  \caption{Real part of the spatial structure of the adjoint mode $1$ obtained with the $k-\omega$ model of Wilcox.} \label{figadjoint}
\end{figure}
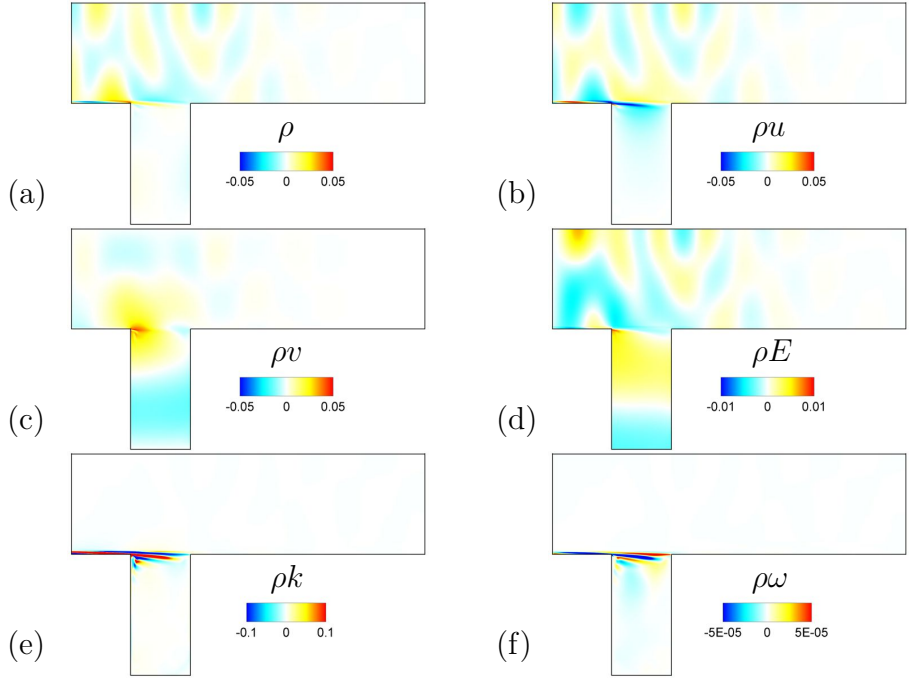

As detailed in \S \ref{sec1}, the resolution of the adjoint problem in Eq. (\ref{adjoint}) is necessary to obtain the sensitivity gradients. We use the discrete inner-product defined such that :
\begin{equation}
\forall\left(\mathbf{u}\text{,}\mathbf{v}\right) \hspace{0.5cm} <\mathbf{u}\text{,}\mathbf{v} >=\sum_{i,j}{u_{i}^{*}v_{i}\Omega_{ij}}=\mathbf{u}^{*}\mathbf{Q}\mathbf{v},
\end{equation}
where $\mathbf{Q}$ is a diagonal matrix whose terms correspond to the surface of the mesh cells. The spatial structure of the fundamental adjoint mode $\tilde{\mathbf{w}}|_{\mathbf{Q}}$ for the $k-\omega$ model of Wilcox is plotted in Fig.\ref{figadjoint}. Note that the adjoint mode is normalized according to Eq. (\ref{adjoint}). As for the direct modes, turbulent scales and mean field quantities present similar structures. Adjoint modes are mostly located upstream the leading edge of the cavity: direct modes propagate downstream while adjoint modes propagate upstream, which comes from the opposite transport of the perturbations by the baseflow in the direct and adjoint linear operators \cite{Sipp2007}. The structure of the adjoint modes obtained with the Spalart-Allmaras model are similar to those obtained with the $k-\omega$ model of Wilcox. \\

\subsection{Sensitivity analysis}

\subsubsection{Sensitivity gradients to baseflow perturbations $\boldsymbol{\nabla}_{\mathbf{w_{b}}}\lambda$}

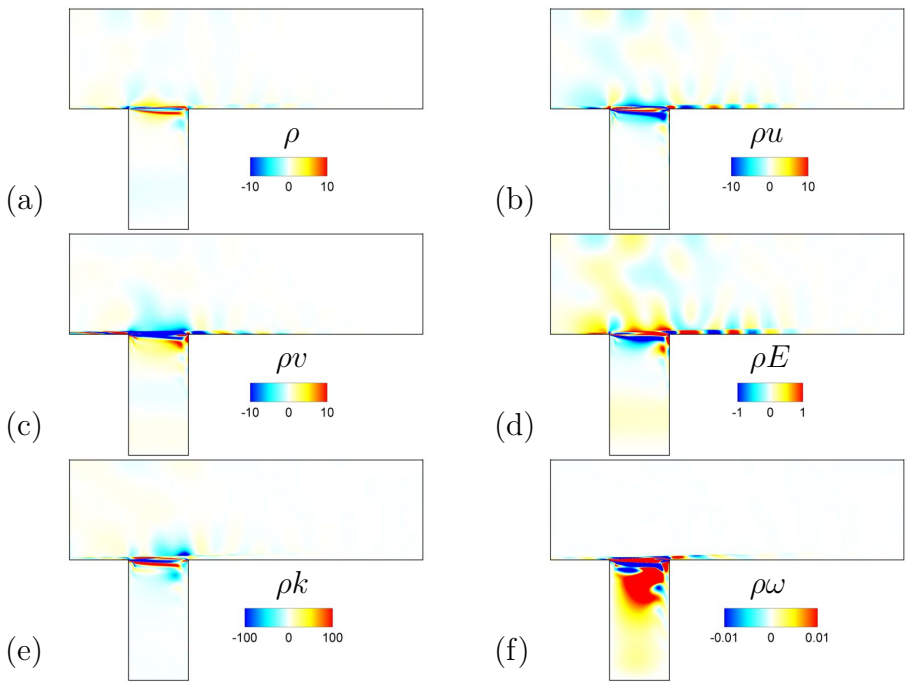
\begin{figure}[h]
  \setlength{\unitlength}{1cm}
  \begin{picture}(10,9)
    \put(0.8,6){\includegraphics*[width=0.35\textwidth]{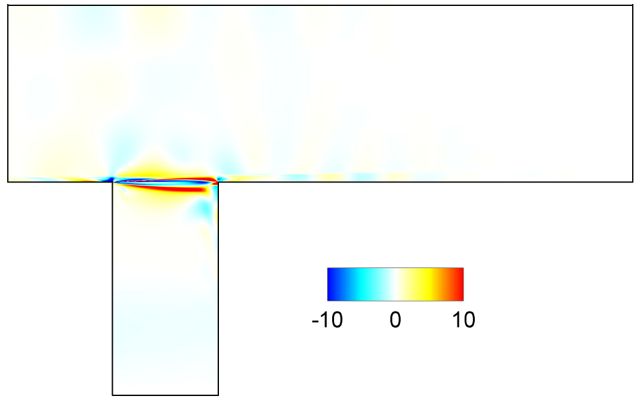}}
    \put(0,6.3){(a)}
    \put(0.8,3){\includegraphics*[width=0.35\textwidth]{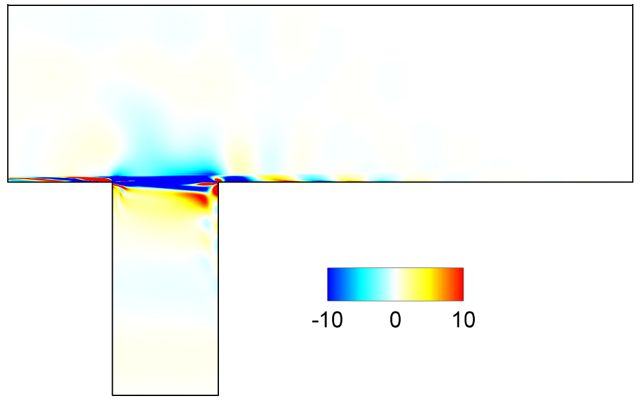}}
    \put(0,3.3){(c)}
    \put(0.8,0){\includegraphics*[width=0.35\textwidth]{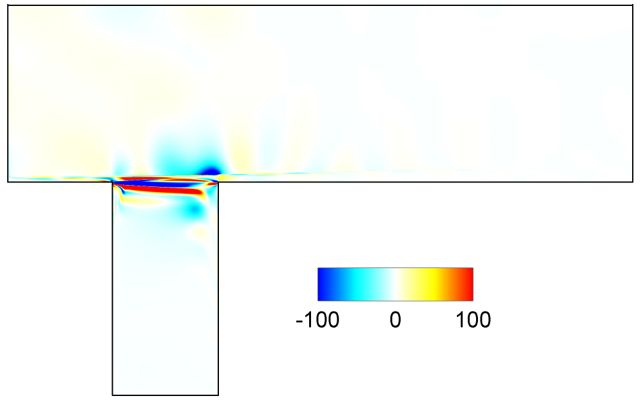}}
    \put(0,0.3){(e)}
    \put(3.7,7.2){$\rho$}
    \put(3.6,4.2){$\rho v$}
    \put(3.6,1.2){$\rho k$}
    \put(7.2,6){\includegraphics*[width=0.35\textwidth]{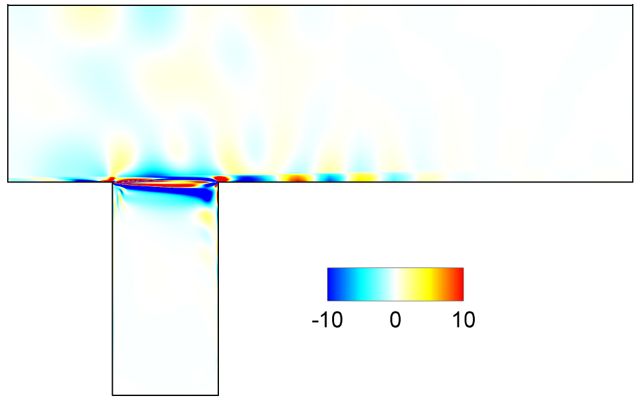}}
    \put(6.5,6.3){(b)}
    \put(7.2,3){\includegraphics*[width=0.35\textwidth]{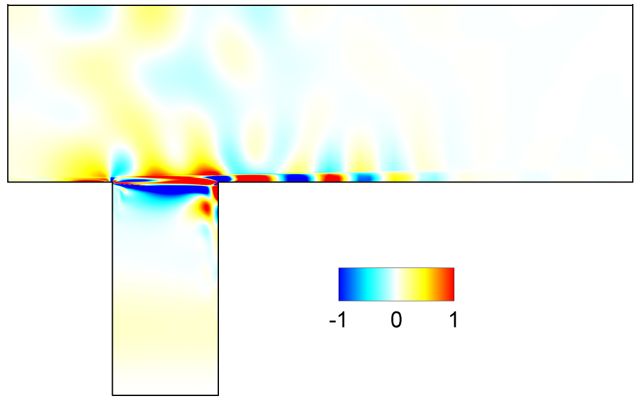}}
    \put(6.5,3.3){(d)}
    \put(7.2,0.0){\includegraphics*[width=0.35\textwidth]{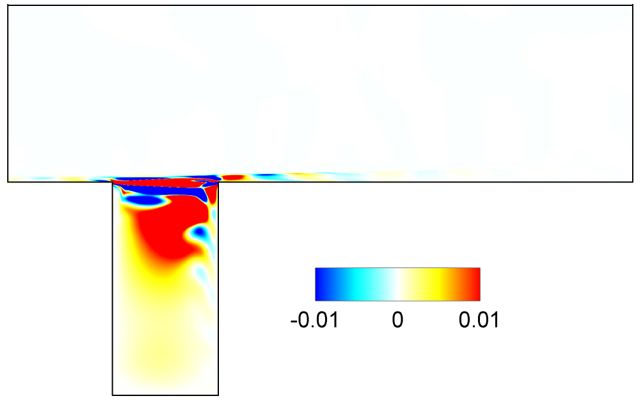}}
    \put(6.5,0.3){(f)}
    \put(9.9,7.2){$\rho u$}
    \put(9.9,4.2){$\rho E$}
    \put(9.9,1.2){$\rho\omega$}
  \end{picture}
  \caption{Sensitivity gradient to baseflow perturbations $\boldsymbol{\nabla}_{\mathbf{w_{b}}}\lambda|_{\mathbf{Q}}$ of mode $1$ obtained with the $k-\omega$ model of Wilcox.}  \label{figgradW}
\end{figure}

Once both direct and adjoint modes are available, we compute the sensitivity gradient to baseflow perturbations $\boldsymbol{\nabla}_{\mathbf{w_{b}}}\lambda|_{\mathbf{Q}}$ as presented in \S \ref{sec1}. As stated by \citet{Marquet2008a}, the sensitivity analysis to baseflow modifications is appropriate to determine which
regions of the baseflow participate to the development of the instabilities. The real part of these fields are  plotted in Fig.\ref{figgradW} for mode $1$ obtained with the $k-\omega$ model of Wilcox and the Roe scheme.  We observe that the eigenvalue is mostly sensitive to perturbations of the baseflow in the mixing layer area which corresponds to the region where Kelvin-Helmholtz instabilities are active.

From a physical point of view, the gradient $\boldsymbol{\nabla}_{\mathbf{w_{b}}}\lambda$ corresponds to the baseflow perturbation that yields the strongest eigenvalue variation \cite{Marquet2008a}. From a numerical point of view, it indicates which areas of the baseflow shall be well captured by the mesh discretization in order to accurately compute the eigenvalues.

\subsubsection{Validation of the gradient $\boldsymbol{\nabla}_{\mathbf{w_{b}}}\lambda$}

\begin{figure} [h!]
  \setlength{\unitlength}{1cm}
  \begin{picture}(8,6)
    \put(0.2,.2){\includegraphics*[width=0.5\textwidth]{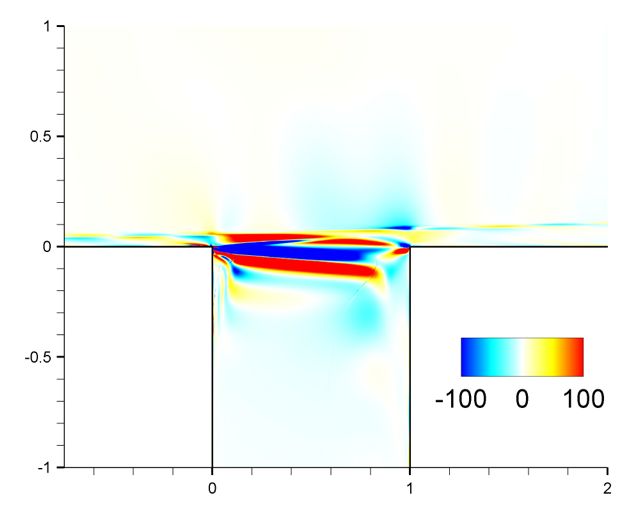}}
    \put(0.1,.2){(a)}
    \put(4,0){\large{$x$}}
    \put(0.1,2.9){\large{$y$}}
    \put(7,0.2){\includegraphics*[width=0.5\textwidth]{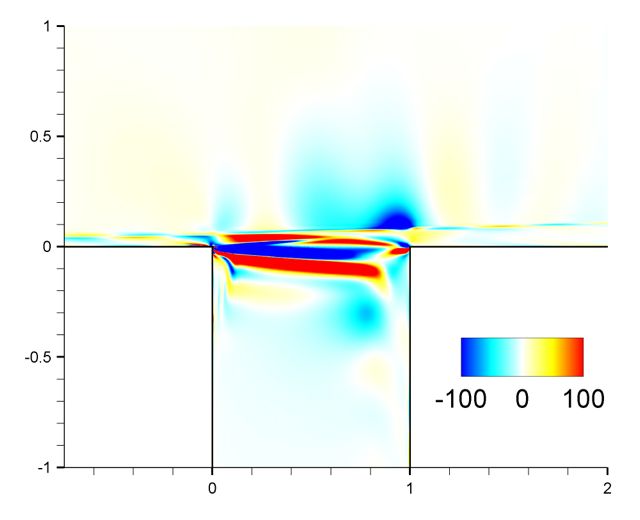}}
    \put(7,0.2){(b)}
    \put(10.6,0){\large{$x$}}
  \end{picture}
  \caption{Comparison of the $\rho k$ component of the sensitivity gradient to baseflow perturbations $\boldsymbol{\nabla}_{\mathbf{w_{b}}}\lambda|_{\mathbf{Q}}$ obtained with (a): the modififed elsA code, and (b): the fully discrete method.}
\label{gradopt}
\end{figure}
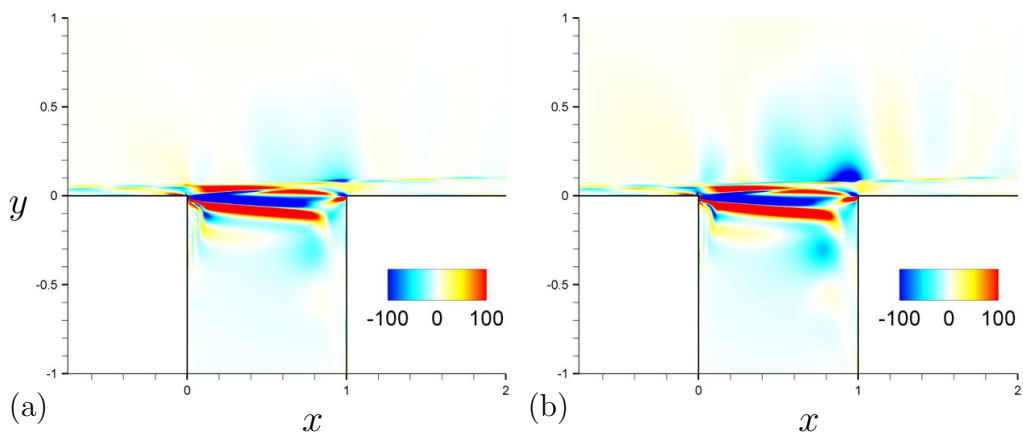

In order to validate our gradients, we first compare the results obtained in the fully discrete approach with those obtained with the modified elsA code. Both methods lead to similar gradient fields but with slightly different amplitudes suggesting the equivalence of both methods. As an example, we plot in Fig.\ref{gradopt} the real part of the $\rho k$ component obtained using both methods.

To validate the gradient,  we then compared, for an arbitrary direction $\mathbf{w_{1}}$, the eigenvalue variation obtained with the gradient, $\delta\lambda=\left<\boldsymbol{\nabla}_{\mathbf{w_{b}}}\lambda\text{,}\mathbf{w_{1}}\right> $, to the eigenvalue variation obtained with a finite difference method,  $\delta\lambda_{1} = \dfrac{1}{\beta}\left[ \lambda\left(\mathbf{w_{b}}+\beta \mathbf{w_{1}}\right) - \lambda\left(\mathbf{w_{b}}\right)\right]$. For this, the Jacobians $\mathbf{J}|_{\mathbf{w_{b}}+\beta\mathbf{w_{1}}}$ and $\mathbf{J}|_{\mathbf{w_{b}}}$  are extracted and their spectrum computed.

Note that the discrete evaluation of $\delta\lambda_{1}$ is a complex issue in itself: the baseflow perturbation $\beta\mathbf{w_{1}}$ shall be small compared to the baseflow although its various components may scale differently from one another. In order to ease the computation of $\delta\lambda_{1}$, we can use the fact that by linearity the full perturbation effect of $\beta \mathbf{w_{1}}$ can be computed from the contributions of its various components separately. We chose $\mathbf{w_{1}}=\mathbf{w_{b}}$ along with a small value for $\beta$ and restricted $\mathbf{w_{1}}$ to each conservative variable independently, so that we perturb each quantity on the full domain one at a time.

We summarized in Table \ref{tablegrad} the relative difference $\dfrac{\left|\delta\lambda_{1}-\delta\lambda\right|}{\left|\delta\lambda\right|}$ between both eigenvalue variation prediction. We observe that the gradient is correctly evaluated up to within $3\%$ for the Spalart-Allmaras model and $ 0.4\%$ for the $k-\omega$ model for each perturbation vector.

\begin{table}[h]
\centering       
\setlength{\tabcolsep}{1pt}
\begin{tabular}{c|c|c|c|c|c|c|c}
 $\mathbf{w_{1}}$           & $\rho$ &  $\rho u$  &  $\rho v$  & $\rho E$   & $\rho k$  & $\rho\omega$  &  $\rho\tilde{\nu}$   \\
\hline
Spalart-Allmaras      &   $0.01$    & $0.001$ & $0.02$ & $0.01$ & . & . & $0.03$ \\
$k-\omega $ &  $0.003$    & $0.001$ & $0.004$ & $0.002$ & $0.0004$ & $0.002$ & .
\end{tabular}
\caption{\label{tablegrad} Relative difference between eigenvalue variation predicted with the sensitivity gradient and a discrete evaluation. }
\end{table}

This validation process also enabled us to determine accurately the best set of $\epsilon_{2}$ values in Eqs. (\ref{fulldiscHessian}) and (\ref{fulldiscHessian2}). The perturbation parameter was fixed with $\epsilon_{2}=\epsilon_{m_{2}}\left(\left|w\right|+1\right)$ where $\left|w\right|$ is the local baseflow value. The best set of $\epsilon_{m_{2}}$ is obtained using different values of $\epsilon_{m_{2}}$ adapted to each conservative variable: these values are summarized in Table \ref{tableeps}.

\begin{table}[h]
\centering       
\setlength{\tabcolsep}{1pt}
\begin{tabular}{c|c|c|c|c|c|c|c}
 Model           & $\rho$ &  $\rho u$  &  $\rho v$  & $\rho E$   & $\rho k$  & $\rho\omega$  &  $\rho\tilde{\nu}$   \\
\hline
Spalart-Allmaras      &   $10^{-4}$    & $10^{-4}$ & $10^{-5}$ & $10^{-3}$ & . & . & $10^{-6}$ \\
$k-\omega $ &  $10^{-5}$    & $10^{-5}$ & $10^{-5}$ & $10^{-3}$ & $10^{-6}$ & $10^{-4}$ & .
\end{tabular}
\caption{\label{tableeps} Linearization parameter $\epsilon_{m_{2}}$ used for the computation of the sensitivity gradient to baseflow perturbations  $\boldsymbol{\nabla}_{\mathbf{w_{b}}}\lambda$. }
\end{table}

The sensitivity gradient $\boldsymbol{\nabla}_{\mathbf{w_{b}}}\lambda$ indicates where and how a baseflow perturbation would affect the unstable eigenvalues and consists in a first step in view of steady control. The question is then how to generate this baseflow perturbation with a meaningful control device, which we consider here as a steady volumic source term in the governing equations. It is thus of interest to consider the sensitivity gradient to a steady force $\boldsymbol{\nabla}_{\mathbf{f}}\lambda$.

\subsubsection{Steady control}

Sensitivity gradients of the unstable eigenvalue to a steady force $\boldsymbol{\nabla}_{\mathbf{f}}\lambda$ are readily obtained from the sensitivity gradients to baseflow perturbations $\boldsymbol{\nabla}_{\mathbf{w_{b}}}\lambda$ through Eq. (\ref{gradient2}). This gradient indicates locations in the flow were a steady force $\boldsymbol{\delta }\mathbf{f}$ could lead to stabilization/destabilization of the unstable modes \cite{Marquet2008a}.

Rather than looking at the gradient fields $\boldsymbol{\nabla}_{\mathbf{f}}\lambda|_{\mathbf{Q}}$, we propose to consider the impact of an infinitesimal control cylinder located at $\left( x\text{,}y\right)$ on the eigenvalue variation $\delta\lambda$ using Eq. (\ref{gradient2}). Similarly to \citet{Marquet2008a}, the local force $\mathbf{f}_{xy}$ that the cylinder exerts on the fluid is taken as a first approximation as proportional and opposite to the drag experienced by the cylinder placed in the baseflow:
\begin{equation}
\mathbf{f}_{xy}\propto-\mathbf{U}_{xy}/ \Omega_{xy}
\end{equation}
where $\Omega_{xy}$ corresponds to the volume of the cell located at $\left(x\text{,}y\right)$.

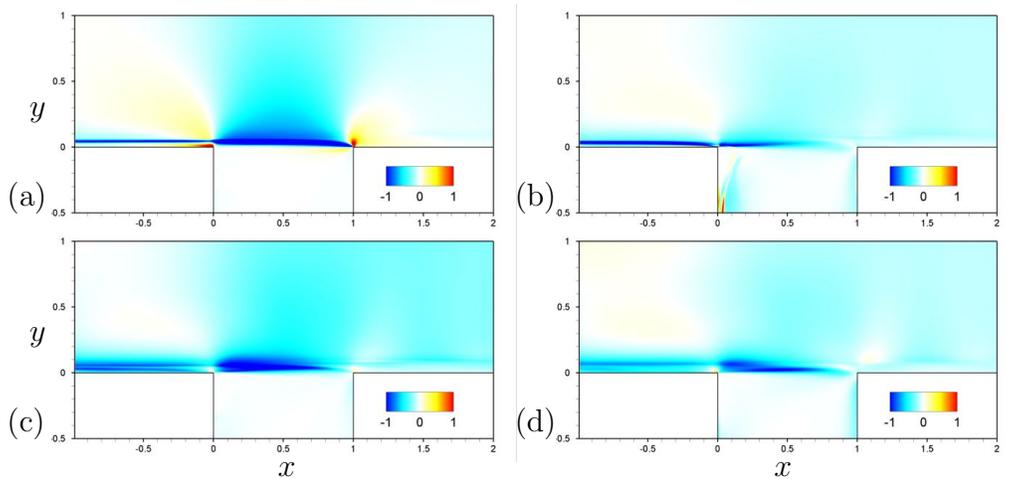
\begin{figure} [h!]
  \setlength{\unitlength}{1cm}
  \begin{picture}(10,7)
    \put(0.6,3.2){\includegraphics*[width=0.45\textwidth]{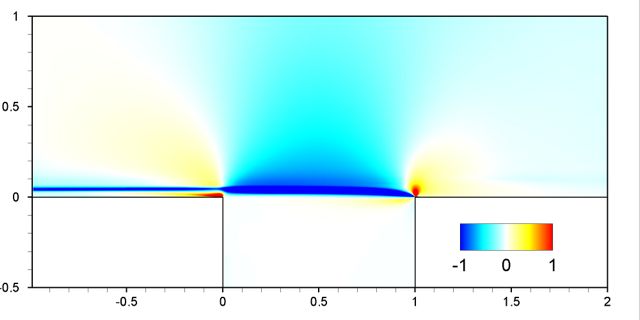}}
    \put(0.0,3.6){(a)}
    \put(0.3,4.8){$y$}
    \put(7.3,3.2){\includegraphics*[width=0.45\textwidth]{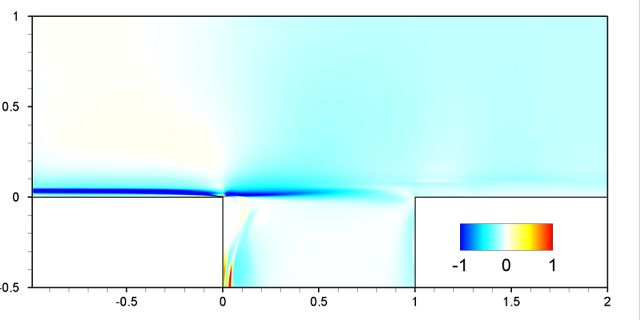}}
    \put(6.75,3.6){(b)}
    \put(0.6,0.2){\includegraphics*[width=0.45\textwidth]{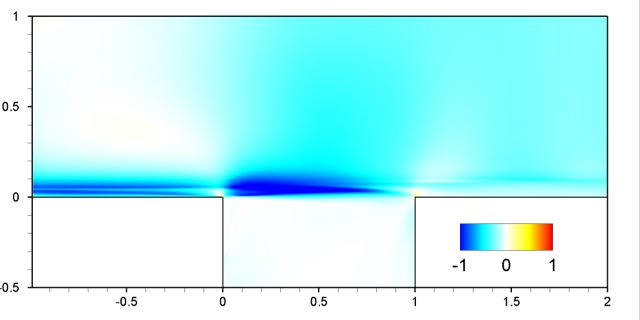}}
    \put(0.0,0.6){(c)}
    \put(3.6,0){$x$}
    \put(0.3,1.8){$y$}
    \put(7.3,0.2){\includegraphics*[width=0.45\textwidth]{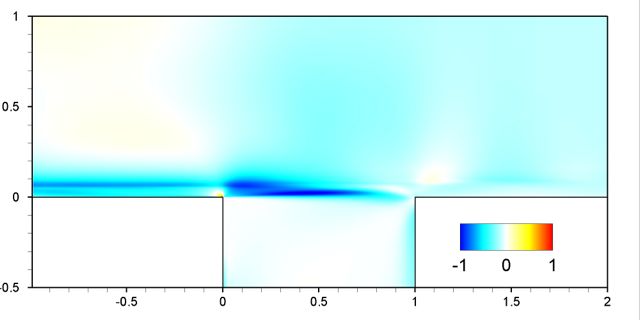}}
    \put(6.75,0.6){(d)}
    \put(10.2,0){$x$}
  \end{picture}
  \caption{Variation of the eigenvalue growth rate $\delta\sigma_{xy} / \left\|\lambda\right\| $ due to the presence of a control cylinder at $\left( x\text{,}y\right)$ (mode $1$). Blue regions indicate that the amplification rate is lowered while increased in the red regions. (a): $k-\omega$ model of Wilcox. (b): Spalart-Allmaras model. (c): elsA optimization code. (d): Uncoupled equations.} \label{finalFig}
\end{figure}

The choice of such a simple model to represent the effort of the control cylinder is motivated by the fact that we are mainly interested in the direction of the eigenvalue variation (stabilization or destabilization). More sophisticated models can be found in \cite{Marquet2008a,Meliga2012}. Computing this force for each cell location in our mesh, we obtain the eigenvalue variation field $\delta\lambda_{xy}$ which indicates how the eigenvalue is impacted by the presence of an infinitely small control cylinder located at $\left( x\text{,}y\right)$.  The real part of $\delta\lambda_{xy}$ corresponds to the growth rate variation $\delta\sigma_{xy}$ of the mode while its imaginary part refers to its frequency change $\delta\omega_{xy}$. In particular, negative values of $\delta\sigma_{xy}$ indicate that the mode growth rate is decreased when the cylinder is located at $\left( x\text{,}y\right)$, which thus induces a stabilizing effect. On the opposite, if  $\delta\sigma_{xy}$ is positive then the cylinder destabilizes the mode and no control effect shall be observed.

We plot in Fig.\ref{finalFig} the field $\delta\sigma_{xy}$ for the different turbulence models that were studied (the maximum value was set to $1$ in each Figure). We observe that the control maps slightly differ from one modeling to an other. In all cases, we recover a stabilization region in blue near $y=0.05$ that extends upstream and downstream of the leading edge of the cavity. These results are in agreement with the experimental study by \citet{Illy2008} whom configuration was similar. They controlled the flow using a small steady cylinder located at the station $\left(-0.1,y\right)$ with $0\leq y\leq 0.22$. They found a critical region $0.05 < y < 0.12$ in which the cylinder had to be placed to control the flow unsteadiness. A small destabilizing region is also obtained just upstream the leading edge of the cavity. We observe small differences between the elsA optimization code results and our fully discrete approach (Fig.\ref{finalFig}(a)\&(c)) which are likely to be linked to the approximations done in the optimization code.

\section{Concluding remarks}

A fully discrete formalism was introduced to perform a stability analysis of a turbulent compressible flow whom dynamics is modeled using the RANS equations. The discrete equations were linearized using finite differences applied to the evaluation of the Navier-Stokes residual $\cal{R}$. The stability of the flow is assessed by solving the direct and adjoint eigenvalue problems linked to the Jacobian matrix $\mathbf{J}$. In the view of open loop control, the sensitivity gradient of the unstable eigenvalue to baseflow perturbations $\boldsymbol{\nabla}_{\mathbf{w_{b}}}\lambda$ was defined within this discrete formalism. In particular, the computation of the gradient was linked to the computation of the Hessian of the RANS equations. The proposed procedure to compute the gradient with finite differences avoids the tedious analytical linearization of the equations. The method is generic regarding the system of equations (turbulence model, numerical scheme) and the code used for the evaluation of $\cal{R}$ can be used in a black box manner. Finally, the sensitivity gradient to a steady force $\boldsymbol{\nabla}_{\mathbf{f}}\lambda$ was introduced, indicating interesting areas of the flow where a steady force could lead to the stabilization of the unstable eigenvalue.

An explicit storage of matrices strategy was adopted, which allows immediate access to adjoint matrices required for the computation of the sensitivity gradients. Both direct and adjoint problems where solved using direct methods for matrix inversions. This strategy is fast and accurate and exploits the sparsity o the Jacobian matrix. It however remains costly in terms of computational memory. An on-the-fly strategy was hence described to tackle three dimensional configurations.

The method was tested on a turbulent compressible flow in a deep cavity. The flow was found to be unstable, in particular the fundamental frequency of the flow was recovered and several of its harmonics were obtained. We obtained unstable modes with  Kelvin-Helmholtz structure as suggested by the instability mechanism of the flow. The acoustic features of the flow were also captured as we observed acoustic resonance modes excited by Kelvin-Helmholtz instabilities.

The impact of the numerical discretization was investigated and appeared to poorly affect the spectrum of the flow. On the contrary, the choice of turbulence model had a slight impact on the growth rates of the unstable eigenvalues but not on their frequency. Convergence properties of the spectrum with the linearization parameter were analysed. The sensitivity gradients were then computed and the choice of the linearization parameters were described. In particular, the gradients were validated using a discrete evaluation of the eigenvalue variations for both turbulence models. The error in predicting the eigenvalue slope was found to be lower then 5\% suggesting that the gradients were correctly computed. Finally, control maps using a steady cylinder as a means to control the flow were obtained for the different turbulence models. Control maps were observed to slightly differ from one model to another. The flow is mostly receptive near the mixing layer and a stabilization region was found for all the turbulence models tested.


\appendix
\section{} \label{ap1}

The continuous form of the mean field fluxes in Eq. (\ref{eq1split}) of the Navier-Stokes equations are given by:
\begin{equation}
\cal{R}^{\text{c,mf}}=-
\begin{pmatrix}
\rho\mathbf{U} \\
\rho\mathbf{U}\otimes\mathbf{U} + p\mathbf{I} \\
\rho E\mathbf{U}+p\mathbf{U}
\end{pmatrix} \hspace{1cm}
\cal{R}^{\text{d,mf}}=
\begin{pmatrix}
0 \\
\boldsymbol{\tau } + \boldsymbol{\tau_{r}} \\
\boldsymbol{\tau }\mathbf{U}  + \boldsymbol{\tau_{r}}\mathbf{U}-\mathbf{q}-\mathbf{q_{t}}
\end{pmatrix},
\end{equation}
with
\begin{equation} 
p=\rho RT \hspace{0.5cm} \boldsymbol{\tau}=-\dfrac{2}{3}\mu\left(\nabla\cdot\mathbf{U}\right)\mathbf{I} + 2\mu\mathbf{D} \hspace{0.6cm} \mathbf{q}=-\dfrac{c_{p}\mu}{Pr}\nabla T ,
\end{equation}
\begin{equation}
\boldsymbol{\tau_{r}}=-\dfrac{2}{3}\mu_{t}\left(\nabla\cdot\mathbf{U}\right)\mathbf{I}  + 2\mu_{t}\mathbf{D} \hspace{0.5cm} \mathbf{q_{t}}=-\dfrac{c_{p}\mu_{t}}{Pr_{t}}\nabla T ,
\end{equation}
$p$ is the pressure, $R$ the perfect gas constant, $c_{p}$ the heat capacity at constant pressure, $\mu$ the viscosity, $T$ the temperature, $\boldsymbol{\tau} $ the viscous tensor, $\mathbf{q}$ the heat flux, $\mathbf{D}$ and $\mathbf{I}$ the strain and identity tensors respectively, $\mu_{t}$ the eddy viscosity (computed with the chosen turbulence model),  $\boldsymbol{\tau_{r}}$ the Reynolds tensor, $\mathbf{q_{t}}$ the flux of diffusion of turbulent enthalpy, $Pr$ and $Pr_{t}$ the classical and turbulent Prandtl number assumed constants and taken respectively equal to $0.72$ and $0.9$.

The preceding equations were derived using Boussinesq hypothesis, perfect gaz relations and neglecting the turbulent kinetic energy term  $k$ in the energy conservative equation as suggested by dimensional analysis for  high Reynolds number flows. The viscosity is computed using Sutherland's law:
\begin{equation}
\mu=\mu_{s}\sqrt{\dfrac{T}{T_{s}}}\dfrac{1+C_{s}/T_{s}}{1+C_{s}/T},
\end{equation}
using the adimentionalized constants $\mu_{s}=1.59~10^{-6}$,$C_{s}=0.43$ and $T_{s}=1.05$.
The variables  $\mathbf{U},E,k,\omega $ are mass weighted averaged using Favre average whereas the other ones are averaged according to the classical RANS average in time. \\

The $k-\omega$ model of Wilcox \cite{Wilcox1988} introduces the turbulent conservative variables $\mathbf{w}^{\text{tf}}=\left(\rho k\text{, }\rho\omega\right)^{T}$. The turbulent fluxes and source terms are then given by  (constants used are given in Table \ref{table1}):

\begin{equation}
\cal{R^{\textbf{c,tf}}}=-
\begin{pmatrix}
\rho k\mathbf{U}   \\
\rho\omega\mathbf{U}
\end{pmatrix}, \hspace{1.0cm}
\cal{R^{\textbf{d,tf}}}=
\begin{pmatrix}
\left( \mu +\sigma^{*}\mu_{t}\right)\nabla k  \\
\left( \mu +\sigma\mu_{t}\right)\nabla\omega
\end{pmatrix},
\end{equation}
\begin{equation}
\cal{T}=
\begin{pmatrix}
\boldsymbol{\tau_{r}}:\boldsymbol{\nabla}\mathbf{U}-\beta^{*}\rho k\omega \\
 \dfrac{\gamma}{\nu_{t}}\boldsymbol{\tau_{r}}:\boldsymbol{\nabla}\mathbf{U} - \beta\rho\omega^{2}
\end{pmatrix}.
\end{equation} \\

The turbulent eddy viscosity is defined by:
\begin{equation} 
\mu_{t}=\dfrac{\rho k}{\omega} .
\end{equation}

\begin{table}
\centering       
\setlength{\tabcolsep}{1pt}
\begin{tabular}{c|c|c|c|c|c}
$\beta^{*}$&$\beta$&$\sigma^{*}$&$\sigma $&$\gamma$& $K$    \\
\hline
0.09   & 0.075   & 0.5   & 0.5   & $\dfrac{\beta}{\beta^{*}}-\dfrac{\sigma K^{2}}{\sqrt{\beta^{*}}}$& 0.41       \\
\end{tabular}
\caption{\label{table1} Constants used in the $k-\omega$ model of Wilcox.}
\end{table}

The Spalart-Allmaras model \cite{Spalart1992} introduces one turbulent conservative variable $\mathbf{w}^{\text{tf}}=\left(\rho\tilde{\nu}\right)$. The turbulent fluxes and source terms are then given by :

\begin{equation}
 \cal{R^{\textbf{c,tf}}}=-
\begin{pmatrix}
\rho\tilde{\nu}\mathbf{U}
\end{pmatrix} \hspace{1.0 cm}
\cal{R^{\textbf{d,tf}}}=
\begin{pmatrix}
\dfrac{\mu +\rho\tilde{\nu}}{\sigma_{\tilde{\nu}}}\nabla\tilde{\nu}
\end{pmatrix},
\end{equation}
\begin{equation}
\cal{T}=
\begin{pmatrix}
C_{b1}\left( 1-f_{t2}\right)\tilde{S}\rho\tilde{\nu} + \dfrac{C_{b2}}{\sigma}\nabla\rho\tilde{\nu}\cdot\nabla\tilde{\nu}
-\left( C_{w1}f_{w}-\dfrac{C_{b1}}{K^{2}}f_{t2}  \right)\rho\dfrac{\tilde{\nu}^{2}}{\eta^{2}}
\end{pmatrix},
\end{equation}
with, noting $\bar{\omega}$ the module of the vorticity : \\
\newsavebox\equa
\begin{lrbox}{\equa}
\begin{minipage}{0.3\textwidth}
    \begin{eqnarray*}
     \tilde{S}&=&\bar{\omega}+\dfrac{\tilde{\nu}}{K^{2}\eta^{2}}f_{v2}\text{,} \\
      g&=&r+C_{w2}\left( r^{6}-r\right)\text{,}
    \end{eqnarray*}
    \end{minipage}
\end{lrbox}
\newsavebox\equab
\begin{lrbox}{\equab}
\begin{minipage}{0.3\textwidth}
    \begin{eqnarray*}
     f_{v2}&=&1-\dfrac{\chi}{1+\chi f_{v1}} \text{,} \\
      f_{t2}&=&C_{t3}e^{-C_{t4}\chi^{2}} \text{,}
    \end{eqnarray*}
     \end{minipage}
\end{lrbox}
\newsavebox\equac
\begin{lrbox}{\equac}
\begin{minipage}{0.3\textwidth}
    \begin{eqnarray*}
     f_{w}&=&g\left(\dfrac{1+C^{6}_{w3}}{g^{6}+C^{6}_{w3}}\right)^{1/6}\text{,}\\
     r&=&\dfrac{\tilde{\nu}}{\tilde{S}K^{2}\eta^{2}}.
    \end{eqnarray*}
     \end{minipage}
\end{lrbox}
\begin{tabular}{ccc}
 \usebox{\equa} &  \usebox{\equab}  &  \usebox{\equac}
\end{tabular}

The turbulent eddy viscosity is defined by:
\begin{equation}
\mu_{t}=\rho\tilde{\nu}f_{v1},
\end{equation}
with :
\begin{equation}
f_{v1}=\dfrac{\chi^{3}}{\chi^{3}+C_{v1}^{3}}\text{,} \hspace{1cm} \chi=\dfrac{\rho\tilde{\nu}}{\mu}.
\end{equation}
The values of the constants for the Spalart-Allmaras model are given in Table \ref{table2}.

\begin{table}
\centering       
\setlength{\tabcolsep}{1pt}
\begin{tabular}{c|c|c|c|c|c|c|c|c|c}
$C_{b1}$ & $C_{b2}$ & $\sigma$ & $K $ & $C_{w1}$ & $C_{w2}$ & $C_{w3}$  &$C_{v1}$ & $C_{t3}$ & $C_{t4}$ \\
\hline
0.1355& 0.622 & 2/3 &0.41 & $C_{b1}/K^{2}+(1+C_{b2})/\sigma$ & 0.3 &2 &$7.1$ & $1.2$ &$0.5$  \\
\end{tabular}
\caption{\label{table2} Constants used in the Spalart-Allmaras model.}
\end{table}

\section{} \label{ap2}

In the following, unstable mode fluctuations are denotted with $^{\prime}$ to distinguish them from baseflow quantities. The eddy viscosity fluctuation $\mu_{t}^{\prime}$ associated to a given mode for the $k-\omega$ model of Wilcox is defined by:
\begin{equation}
\mu_{t}+\mu_{t}^{\prime}=\dfrac{\left( \rho + \rho^{\prime}\right)\left( \rho k + \left(\rho k\right)^{\prime}\right)}{\left( \rho\omega + \left(\rho\omega\right)^{\prime}\right)}.
\end{equation}
That is to the first order:
\begin{equation}
\mu_{t}^{\prime}=\dfrac{\rho k }{\rho\omega}\rho^{\prime}+\dfrac{\rho}{\rho\omega}\left(\rho k\right)^{\prime}-\dfrac{\rho\rho k }{\left(\rho\omega\right)^{2}}\left(\rho\omega\right)^{\prime}.
\end{equation}
For the Spalart-Allmaras turbulence model we have:
\begin{equation}
\mu_{t}^{\prime}=\mu\left(\dfrac{\mu_{t}}{\mu}\right)^{\prime}+\dfrac{\mu_{t}}{\mu}\mu^{\prime},
\end{equation}
with:
\begin{eqnarray}
\left(\dfrac{\mu_{t}}{\mu}\right)^{\prime}&=&\dfrac{4\chi^{3}\left(\chi^{3}+C^{3}_{v1}\right)-3\chi^{6}}{\left(\chi^{3}+C^{3}_{v1}\right)^{2}}\chi^{\prime}, \\
\chi^{\prime}&=&\dfrac{\left(\rho\tilde{\nu}\right)^{\prime}}{\mu}-\dfrac{\rho\tilde{\nu}}{\mu^{2}}\mu^{\prime}, \\
\mu^{\prime}&=&\dfrac{\mu_{s}}{\sqrt{T_{s}}}\dfrac{1+C_{s}/T_{s}}{1+C_{s}/T}\left[ \dfrac{1}{2\sqrt{T}}
+\dfrac{\sqrt{T}C_{s}}{T^{2}\left(1+C_{s}/T\right)} \right]T^{\prime}, \\
T&=& \dfrac{\gamma -1}{\rho R}\left( \rho E -0.5\rho u^{2} -0.5\rho v^{2}\right),\\
T^{\prime}&=&-\dfrac{T}{\rho}\rho^{\prime}+\dfrac{\gamma -1}{\rho R}\left(\rho E\right)^{\prime} \\ \notag
&+&\dfrac{\gamma -1}{\rho R}\left[
-\dfrac{\rho u}{\rho}\left(\rho u\right)^{\prime}+0.5\dfrac{\left(\rho u\right)^{2}}{\rho^{2}}\rho^{\prime}
-\dfrac{\rho v}{\rho}\left(\rho v\right)^{\prime}+0.5\dfrac{\left(\rho v\right)^{2}}{\rho^{2}}\rho^{\prime}\right].
\end{eqnarray}

\section*{Acknowledgments}
The authors acknowledge Jacques Peter (Computational Fluid Dynamics and Aeroacoustics Department, ONERA) and Olivier Marquet (Fundamental and Experimental Aerodynamics Department, ONERA) for usefull discussions.

\bibliographystyle{elsarticle-num-names}
\bibliography{smallBIB}







\end{document}